\documentclass[10pt, two column, twoside]{IEEEtran}
\usepackage{amssymb}
\usepackage{amsmath}
\usepackage{mathrsfs}
\usepackage{algorithm}
\usepackage{algorithmic}

\usepackage{tikz}
\usetikzlibrary{arrows}
\usepackage{subfigure}
\usepackage{graphicx,booktabs,multirow}

\usepackage{color}

\definecolor{colorhkust}{RGB}{20,43,140}
\definecolor{colortsinghua}{RGB}{116,52,129}
\definecolor{color1}{RGB}{128,0,0}
\definecolor{color2}{HTML}{D0B22B}

\newtheorem{proposition}{Proposition}
\newtheorem{definition}{Definition}
\newtheorem{remark}{Remark}

\newcommand{\bs}{\boldsymbol}

\newcommand{\rev}{\color{black}}

\begin{document}

\title{Low-Rank Matrix Completion for Topological
Interference Management by Riemannian Pursuit}
\author{Yuanming~Shi,~\IEEEmembership{Member,~IEEE,}
        Jun~Zhang,~\IEEEmembership{Senior Member,~IEEE,}
        and~Khaled~B.~Letaief,~\IEEEmembership{Fellow,~IEEE}
 \thanks{Manuscript received xxx; revised xxx; accepted xxx. Date of publication
xxx; date of current version xxx. This work is supported by the Hong Kong Research Grant Council under Grant No. 610113.}
\thanks{Y. Shi is with the School of Information Science and Technology,
ShanghaiTech University, Shanghai, China (e-mail: shiym@shanghaitech.edu.cn).}
\thanks{ J. Zhang is with the Department of Electronic and Computer Engineering,
 Hong Kong University of Science and Technology, Hong Kong (e-mail: eejzhang@ust.hk).}

\thanks{K. B. Letaief is with Hamad bin Khalifa University (e-mail: kletaief@hbku.edu.qa)
and Hong Kong University of Science and Technology (e-mail: eekhaled@ust.hk).}}

\maketitle

\begin{abstract}
In this paper, we present a flexible low-rank matrix completion (LRMC) approach for topological interference management  (TIM)  in the partially connected $K$-user interference channel. No channel state information (CSI) is required at the transmitters except the network topology information. The previous attempt on the TIM problem is mainly based on its equivalence to the index coding problem, but so far only a few index coding problems have been solved. In contrast, in this paper, we {\rev{present an algorithmic}} approach to investigate the achievable degrees-of-freedom (DoFs)  by recasting the TIM problem as an LRMC problem. Unfortunately, the resulting LRMC problem is known to be NP-hard, and the main contribution of this paper is to propose a Riemannian pursuit (RP) framework to detect the rank of the matrix to be recovered by iteratively increasing
the rank. This algorithm solves a sequence of fixed-rank matrix completion problems. To address the convergence issues in the existing fixed-rank optimization methods, the quotient manifold geometry of  the search space of fixed-rank matrices is exploited via Riemannian optimization. By further exploiting the structure of the low-rank matrix varieties, i.e., the closure of the set of fixed-rank matrices, we develop an efficient rank increasing strategy to find good initial points in the procedure of rank pursuit. Simulation results demonstrate that the proposed RP algorithm achieves a faster convergence rate and higher achievable DoFs for the TIM problem compared with the state-of-the-art methods.           
\end{abstract}


\section{{Introduction}}
Network densification with interference  coordination has been recognized as   a promising way to meet the exponentially growing mobile data traffic in  next generation wireless networks \cite{Yuanming_WCMLargeCVX,Gesbert_JSAC10, Yuanming_LargeSOCP2014}. In particular, interference alignment \cite{Jafar_IT2008} has been proposed as a powerful tool to understand the Shannon capacity in various interference-limited scenarios, e.g., the MIMO interference channel \cite{Tse_TIT2014feasibilityIA} and cellular networks \cite{Caire_TIT2015}. Although interference alignment can serve as a linear interference management strategy achieving the optimal DoFs in many scenarios, the overhead of obtaining the required global instantaneous channel state information (CSI) has hindered its practical implementation \cite{Heath_TWC2012IA}.  This motivates numerous research efforts on CSI overhead reduction for interference alignment, e.g., with delayed CSI \cite{Tse_TIT2012completely} and alternating CSI \cite{Jafar_TIT2014alt}. However, the practical applicability of these results remain unclear. More recently, a new proposal has emerged, namely, topological interference management (TIM) \cite{Jafar_TIT2013TIM}, as a promising solution for the partially connected interference channels. It is mainly motivated by the fact that most of the channels in a wireless network are very weak and can be ignored due to the shadowing and pathloss \cite{Jafar_TIT2013TIM,Avestimehr_TIT14_TIM, Yuanming_TSP14SCB}. It thus provides an opportunity to manage interference only based on  topological information rather than the instantaneous CSI.

Specifically, in the TIM problem, we assume that no CSI at the transmitters is available beyond the network topology knowledge, i.e., the connectivity of the wireless network. Due to the practical applicability of such CSI assumption and information theoretic interest, the TIM problem has received tremendous attentions and been investigated in various scenarios with partial connectivity, e.g., the fast fading scenarios \cite{Avestimehr_TIT14_TIM, Avestimehr_arXiv2015rank}, transmitter cooperation \cite{Gesbert_TIT2015TIM} and MIMO interference channels \cite{Jafar_ISIT2014TIM_MIMO}. In particular, {\rev{in a slow fading scenario}}, by establishing the connection between the wireless TIM problem and  the wired index coding problem, efficient capacity and DoF analysis was provided in \cite{Jafar_TIT2013TIM} based on the existing results from index coding problems. However, the index coding problem itself is an open problem, and thus the existing solutions are  only valid for some special cases. For general network topologies in the wireless TIM problem, the  optimal DoF is still unknown. {\rev{In a fast fading scenario, a matrix rank-loss approach based on matroid and graph theories was presented in \cite{Avestimehr_arXiv2015rank} to characterize the symmetric DoF for a class of TIM problems.}}

In this paper, we will present an algorithmic approach to evaluate the achievable DoFs in the TIM problem for general partially connected interference channels. It is achieved by recasting the original TIM problem as a low rank matrix completion (LRMC) problem \cite{Candes_2009exactMC}. Then the minimum number of channel uses for interference-free data transmission will be equal to the minimum rank of the matrix in the associated LRMC problem. This approach has recently been applied to solve the linear index coding problem over the finite field \cite{Hassibi_2014matrix} and the wireless TIM problem with symmetric DoFs \cite{Hassibi_TIA2014,Yuanming_ISIT2015TIA}.  We shall extend the previous results on the symmetric DoF case with single data transmission for each user \cite{Hassibi_TIA2014,Yuanming_ISIT2015TIA} to any achievable DoF region.  The presented LRMC approach will serve as a flexible way to maximize the achievable DoFs for any network topology, thereby providing insights on the TIM problem for general network topologies that are not yet available in theory.

Unfortunately, the resulting LRMC problem is NP-hard due to the non-convex rank objective. Although the widely used nuclear norm based convex relaxation provides an effective way to solve the LRMC problem with polynomial time complexity and optimality guarantees with well structured affine constraints \cite{Candes_2009exactMC}, it is inapplicable to our problem as it always returns a full rank solution \cite{Hassibi_TIA2014}. Another category of algorithms is based on alternating minimization \cite{Wotao_2012solvingLR,Jain_2013lowAltmin} by recasting the original LRMC problem as a fixed-rank optimization problem. Although the optimality can be guaranteed with standard assumptions (e.g.,  the original data matrix should be incoherent \cite{Candes_2009exactMC}), the existing fixed-rank methods may converge slowly \cite{Vandereycken2013low,Nicolas_2011rtrmc} and require the optimal rank of the matrix as a prior information \cite{Vandereycken_ICML2014RiemanMatrixRec}.

\subsection{Contributions}
{\rev{We present a low-rank matrix completion approach to maximize the
achievable DoFs for the TIM problem. In particular, we extend the results
 in \cite{Yuanming_ISIT2015TIA,Hassibi_TIA2014} for the symmetric DoF with
single data transmission for each user to any DoF region}}. 
To address the limitations of existing fixed-rank approaches, we propose a Riemannian pursuit (RP) algorithm to solve the LRMC problem for the TIM problem. This is achieved by iteratively increasing the rank of the matrix to be recovered. In particular, the developed RP algorithm possesses the following properties:
\begin{itemize}
\item We can efficiently solve the fixed-rank optimization problems to address the convergence issues in the existing fixed-rank methods;
\item We design an efficient rank increasing strategy to find a good initial point in the next iteration for rank pursuit. 
\end{itemize}

In the proposed RP framework, by exploiting the Riemannian quotient manifold geometry of the search space of fixed-rank matrices via low-rank matrix factorization \cite{Nicolas_2011rtrmc, Mishra_2014fixedrank,Mishra_2014r3mc,Mishra_RP2014}, the nonlinear conjugate gradient (a first-order method with superlinear convergence rate endowed with a good Riemannian metric \cite{Mishra_2014r3mc,Mishra_RP2014}) and trust-region (a second-order method with quadratic convergence rate \cite{Absil_2007trustregion}) based Riemannian optimization algorithms \cite{Absil_2009optimizationonManifolds} are developed to solve the smooth fixed-rank optimization problems. These algorithms can achieve faster convergence rates and higher precision solutions compared with the existing fixed-rank methods, such as the alternating minimization method \cite{Wotao_2012solvingLR,Jain_2013lowAltmin} and the embedded manifold based Riemannian optimization algorithm \cite{Vandereycken2013low}. Furthermore, by exploiting the structures of low-rank matrix varieties \cite{Vandereycken_ICML2014RiemanMatrixRec,Vandereycken_NOLTA2014,Yuanming_ISIT2015TIA}, i.e., the closure of the set of fixed-rank matrices, an efficient rank increasing strategy is proposed to find a high quality initial point and to guarantee that the objective decreases monotonically in the procedure of rank pursuit.

In summary, the major contributions of the paper are as follows:
\begin{enumerate}
\item A Riemannian pursuit  framework is proposed to solve the resulting LRMC problem by solving a sequence of fixed-rank optimization problems with an efficient rank increasing strategy.   
\item To address the convergence issues in the existing fixed-rank based methods, we present a versatile Riemannian optimization framework by exploiting the quotient manifold geometry of the fixed-rank matrices and the least-squares structure of the cost function \cite{Mishra_2014r3mc} as well as the second-order information of the problem.

\item A novel rank increasing strategy is proposed, which considers intrinsic manifold structures in the developed Riemannian optimization algorithms. In particular, by exploiting the structures of low-rank varieties, we extend the results in \cite{Vandereycken_ICML2014RiemanMatrixRec,Yuanming_ISIT2015TIA} for the embedded manifold to the framework of the quotient manifold.   
    
\end{enumerate} 

Simulation results will demonstrate the superiority of the proposed RP algorithms
with  faster convergence rates and the capability of automatic rank detection
compared with the existing fixed-rank optimization algorithms to maximize
the achievable DoFs for the TIM problem.

\subsection{Organization}
The remainder of the paper is organized as follows. Section {\ref{sysmod}} presents the system model and problem formulations. In Section {\ref{lrmc}}, the low-rank matrix completion approach with Riemannian pursuit is developed. The Riemannian optimization algorithms  are developed in Section {\ref{Manopt}}. The rank increasing strategy is presented in Section {\ref{rankinc}}. Numerical results will be demonstrated in Section {\ref{simres}}. Finally, conclusions and discussions are presented in Section {\ref{confw}}. The derivations
of the Riemannian optimization related ingredients are diverted to
the appendix.  

\section{System Model and Problem Statement}
\label{sysmod}
\subsection{Channel Model}
Consider the topological interference management (TIM) problem in the partially connected $K$-user
interference channel with $K$ single-antenna transmitters
and $K$ single-antenna receivers \cite{Jafar_TIT2013TIM}. Specifically, let
$\mathcal{V}$ be the index set of the connected transceiver pairs such that
$(i,j)\in\mathcal{V}$ representing the $i$-th receiver is connected to the
$j$-th transmitter. That is, the channel propagation coefficients 
belonging to the set $\mathcal{V}$ are nonzero and are set to be zeros otherwise. Each transmitter $j$ wishes to send a message $W_j$ to its corresponding receiver $j$. Here, $W_j$ is uniformly chosen in the corresponding message set $\mathcal{W}_j$.

Each transmitter $j$ encodes its message $W_j$ into a vector ${\bf{x}}_j\in\mathbb{C}^{N}$
of length $N$ and transmits the signal over $N$ time slots. Therefore, the
input-output relationship is given by
\setlength\arraycolsep{1pt}
\begin{eqnarray}
\label{inout}
{\bf{y}}_i={\bf{H}}^{[ii]}{\bf{x}}_i+\sum_{(i,j)\in\mathcal{V}, i\ne j}{\bf{H}}^{[ij]}{\bf{x}}_{j}+{\bf{n}}_i,
\forall i=1,\dots, K,
\end{eqnarray} 
where ${\bf{n}}_i\sim\mathcal{CN}({\bf{0}}, {\bf{I}}_N)$ and ${\bf{y}}_i\in\mathbb{C}^{N}$ are the additive isotropic white Gaussian noise and received signal at receiver $i$, respectively; ${\bf{H}}^{[ij]}={\rm{diag}}\{H_{ij}\}={H}_{ij}{\bf{I}}_N$ is an $N\times N$ diagonal
matrix with  $H_{ij}\in\mathbb{C}$ as the  channel
coefficient between transmitter $j$ and receiver $i$ in the considered block.
We consider the block fading channel model, and thus the channel stays constant during the $N$ time slots, i.e., all the diagonal entries in
${\bf{H}}^{[ij]}$ are  the same. {\rev{The matrix representation for the channel coefficients in (\ref{inout}) is mainly for the comparison of different channel models to establish the interference alignment conditions, which will be explained in Section {\ref{tim_intro}}.
In this paper, following the TIM setting \cite{Jafar_TIT2013TIM},
we assume that only the network topology information $\mathcal{V}$ is available
at transmitters.}} Furthermore, each transmitter has
an average power constraint, i.e.,  ${1\over{N}}\mathbb{E}[\|{\bf{x}}_i\|^2]\le
\rho$ with $\rho>0$ as the maximum transmit power.  

\subsection{Achievable Rates and DoF}

We assume that transmitters $1,2,\dots, K$ have independent messages $W_1,W_2,\dots,
W_K$ intended for receivers $1,2,\dots, K$, respectively. The rate tuple
$(R_1,R_2,\dots, R_K)$ with $R_i={\log|\mathcal{W}_i|\over{N}}$ is achievable if there exists
an encoding and decoding scheme such that the probability of decoding error for all
the messages can be  made arbitrarily small simultaneously as the codewords
length $N$ approaches infinity \cite{Cover2012elements}. 

The degrees of freedom (DoF) in the partially connected $K$-user interference channel is defined as \cite{Jafar_TIT2013TIM,Jafar_IT2008} 
\begin{eqnarray}
d_i=\limsup_{\rho\rightarrow\infty}{{R_i}\over{\log(\rho)}}, \forall i.
\end{eqnarray}
The DoF region $\mathcal{D}$ is defined as the closure of the set of achievable
DoF tuples. In particular, the symmetric DoF $d_{\textrm{sym}}$ is the highest value
$d_0$, such that the DoF allocation $d_i=d_0, \forall i$, is inside the DoF
region. This is given by \cite{Jafar_TIT2013TIM}
\begin{eqnarray}
d_{\textrm{sym}}=\limsup_{{\rho}\rightarrow\infty}\left[\sup\nolimits_{(R_{\textrm{sym}},\dots,
R_{\textrm{sym}})\in\mathcal{D}}{R_{\textrm{sym}}\over{\log (\rho)}}\right].
\end{eqnarray}
In this paper, we choose the DoF as the performance metric and design the corresponding linear interference management strategies to maximize the achievable DoFs \cite{Jafar_TIT2013TIM,Tse_TIT2014feasibilityIA}. 

\subsection{Topological Interference Management}
\label{tim_intro}
Linear schemes become  particular interesting for interference management due to their low-complexity and the DoF optimality in many scenarios \cite{Jafar_TIT2013TIM,Jafar_IT2008,Tse_TIT2014feasibilityIA}. We thus restrict the class of interference
management strategies to linear schemes to maximize the achievable DoFs  as the signal-to-noise ratio (SNR) approaches infinity. Specifically, for message $W_j$, let ${\bf{V}}_{j}\in\mathbb{C}^{N\times M_j}$ and ${\bf{U}}_i\in\mathbb{C}^{N\times
M_i}$ be the precoding matrix at transmitter $j$ and the receiver combining matrix
at receiver $i$, respectively. Assume that each message
$W_j$ is split into $M_j$ independent scalar data streams, denoted as ${\bf{s}}_j=[s_1(W_j),
s_2(W_j),\dots, s_{M_j}(W_j)]^{T}\in\mathbb{C}^{M_j}$. And $s_m(W_j)$'s are independent Gaussian codebooks, each of which carries
one symbol and is transmitted along the column vectors of the precoding matrix
${\bf{V}}_j$. Therefore, over the $N$ channel uses, the input-output relationship
(\ref{inout}) is rewritten as
\begin{eqnarray}
\label{inout1}
{\bf{y}}_i={\bf{H}}^{[ii]}{\bf{V}}_i{\bf{s}}_i+\sum_{(i,j)\in\mathcal{V},
i\ne j}{\bf{H}}^{[ij]}{\bf{V}}_j{\bf{s}}_{j}+{\bf{n}}_i,
\forall i.
\end{eqnarray}     

In the regime of asymptotically high SNR, to accomplish decoding, we impose
the constraints that, at each receiver $i$, the desired signal space  ${\bf{H}}^{[ii]}{\bf{V}}_i$
is complementary to the interference space $\sum_{(i,j)\in\mathcal{V}, i\ne
j}{\bf{H}}^{[ij]}{\bf{V}}_j$. That is, after projecting the received signal
vector ${\bf{y}}_i$ onto the space  ${\bf{U}}_i$, the interference terms
should be aligned and then cancelled while the desired signal should be preserved \cite{Tse_TIT2014feasibilityIA,Jafar_TIT2011distributed,Jafar_IT2008},
i.e.,
\begin{eqnarray}
\label{IA1}
{\bf{U}}_i^{\sf{H}}{\bf{H}}^{[ij]}{\bf{V}}_j&=&{\bf{0}}, \forall i\ne j,
(i,j)\in\mathcal{V},\\
\label{IA2}
{\det}\left({\bf{U}}_i^{\sf{H}}{\bf{H}}^{[ii]}{\bf{V}}_i\right)&\ne& 0, 
\forall
i.
\end{eqnarray}
If conditions (\ref{IA1}) and (\ref{IA2}) are satisfied, the parallel
interference-free channels can be obtained over $N$ channel uses. Therefore,
the DoF of $M_i/N$ is achieved for  message $W_i$. However, this requires instantaneous CSI and  its acquisition is challenging in dense networks with a large number of transceiver pairs \cite{Heath_TWC2012IA,Jafar_TIT2013TIM}. 

Observe that the channel
matrix ${\bf{H}}^{[ij]}$ equals  $H_{ij}{\bf{I}}_N$ for the constant
channel over the $N$ channel uses. The conditions  (\ref{IA1}) and (\ref{IA2})
can be rewritten as the following channel independent conditions:
\begin{eqnarray}
\label{TIM1}
{\bf{U}}_i^{\sf{H}}{\bf{V}}_j&=&{\bf{0}}, \forall i\ne j,
(i,j)\in\mathcal{V},\\
\label{TIM2}
{\det}\left({\bf{U}}_i^{\sf{H}}{\bf{V}}_i\right)&\ne& 0, 
\forall
i.
\end{eqnarray}
Therefore, we can design
the transceivers ${\bf{U}}_i$'s and ${\bf{V}}_j$'s only based on the knowledge
of the network topology without requiring the instantaneous CSI. This is fundamentally different from the conventional interference alignment approach \cite{Tse_TIT2014feasibilityIA,Jafar_IT2008, Luo_TSP2012IAMIMO}, in which the global instantaneous CSI is required. In contrast, the channel independent topological interference management conditions (\ref{TIM1}) and (\ref{TIM2}) make the corresponding interference management approach much more practical. 

{\rev{\begin{remark}
In this paper, we consider the block fading channel model to capture
the channel coherence phenomenon in a slow fading scenario. Specifically, we assume
that channel gains stay constant over $N$ time slots such that the effective
channel matrix ${\bf{H}}^{[ij]}$ is a diagonal matrix with identical diagonal
entries, which plays a key role to yield the channel independent interference
alignment conditions (\ref{TIM1}) and (\ref{TIM2}). This further motives the low-rank matrix
completion approach in Section {\ref{lrmc}}. However, in a fast fading scenario, i.e.,
the channel gains change at each time instant, the  approaches presented in
this paper may not be applicable, and other approaches (e.g., the rank-loss approach \cite{Avestimehr_arXiv2015rank})
are required.
\end{remark}}}

The problem of studying the DoFs in the partially connected interference channels based on the network topology information is  known as the \emph{topological interference management} (TIM) problem \cite{Jafar_TIT2013TIM,Avestimehr_TIT14_TIM,Gesbert_ISIT2014TIMCoMP}. Most of the existing works on the TIM problem are trying to establish the topology conditions under which the desired DoF is achievable based on graph theory \cite{Avestimehr_TIT14_TIM,Gesbert_ISIT2014TIMCoMP}, or applying the existing results from the index coding problem \cite{Jafar_TIT2013TIM}. In contrast, in this paper, by generalizing the preliminary results in \cite{Hassibi_TIA2014,Yuanming_ISIT2015TIA} for the case of single data stream transmission, we present a novel approach based on the low-rank matrix completion \cite{Candes_2009exactMC} to solve the TIM problem based on conditions (\ref{TIM1}) and (\ref{TIM2}) for arbitrary network topologies with arbitrary number of data streams. Furthermore, novel algorithms will be developed based on Riemannian optimization techniques \cite{Absil_2009optimizationonManifolds} to solve the resulting NP-hard LRMC problem.

\section{Low-Rank Matrix Completion for Topological Interference
Management via Riemannian Pursuit }
\label{lrmc}
In this section, we present a low-rank matrix completion approach to solve the TIM problem, i.e., finding the minimum channel uses $N$ such that the interference alignment conditions (\ref{TIM1}) and (\ref{TIM2}) are feasible. Specifically, define ${\bf{X}}_{ij}={\bf{U}}_i^{\sf{H}}{\bf{V}}_j\in\mathbb{C}^{M_i\times
M_j}$. Then, conditions (\ref{TIM1}) and (\ref{TIM2}) can be rewritten
as
\begin{eqnarray}
\label{TIM}
\mathcal{P}_{\Omega}({\bf{X}})={\bf{I}}_M,
\end{eqnarray}
where ${\bf{X}}=[{\bf{X}}_{ij}]\in\mathbb{C}^{M\times M}$ with $M=\sum_{i}M_i$,
${\bf{I}}_M$ is the $M\times M$ identity matrix, and $\mathcal{P}_{\Omega}:\mathbb{R}^{M\times
M}\rightarrow\mathbb{R}^{M\times M}$ is the orthogonal projection operator
onto the subspace of matrices which vanish outside  $\Omega$ such that the
$(i,j)$-th component of $\mathcal{P}_{\Omega}({\bf{X}})$ equals to $X_{ij}$
if $(i,j)\in\Omega$ and zero otherwise. Here, the set $\Omega$ is defined
as $\Omega=\{\mathcal{G}_i\times\mathcal{G}_j, (i,j)\in\mathcal{V}\}$, where $\mathcal{G}_i=\{\sum_{k=1}^{i-1}M_k+1,\dots, \sum_{k=1}^{i}M_k\}$. {\rev{For example, given the network topology adjacency matrix $\mathcal{V}=\{(1,1), (1,2), (2,2)\}$ and $M_1=M_2=2$, the set $\Omega$ is given as $\Omega=\{(1,1), (1,2), (2,1), (2,2), (1,3),
(1,4), (2,3), (2,4), (3,3),\\ (3,4), (4,3), (4,4)\}$.}} {\rev{To yield a nontrivial solution, we assume that $N\le M$. As ${\bf{X}}=[{\bf{U}}_i^{\sf{H}}{\bf{V}}_j]={\bf{U}}^{\sf{H}}{\bf{V}}\in\mathbb{C}^{M\times
M}$ with ${\bf{U}}=[{\bf{U}}_1, \dots, {\bf{U}}_K]^{\sf{H}}\in\mathbb{C}^{M\times
N}$, ${\bf{V}}=[{\bf{V}}_1, \dots, {\bf{V}}_K]\in\mathbb{C}^{N\times M}$, we have ${\rm{rank}}({\bf{X}})=N$.}} 

{\rev{\begin{remark}
To assist numerical algorithm design, we specify ${\bf{U}}_i^{\sf{H}}{\bf{V}}_i={\bf{I}}, \forall i$ for condition (\ref{TIM2}) to recover the desired signal. Specifically, for the desired message $W_i$, as ${\bf{U}}_i^{\sf{H}}{\bf{V}}_i$ is invertible, by projecting ${\bf{y}}_i$ onto the ${\bf{U}}_i$ space, we have 
\begin{eqnarray}
\tilde{\bf{y}}_i&=&{1\over{H_{ii}}}\left[{\bf{U}}_i^{\sf{H}}{\bf{V}}_i\right]^{-1}{\bf{U}}_i^{\sf{H}}{\bf{y}}_i\\
&=&{1\over{H_{ii}}}\left[{\bf{U}}_i^{\sf{H}}{\bf{V}}_i\right]^{-1}\left(H_{ii}{\bf{U}}_i^{\sf{H}}{\bf{V}}_i{\bf{s}}_i+{\bf{U}}_i^{\sf{H}}{\bf{n}}_i\right)\\
&=& {\bf{s}}_i+{1\over{H_{ii}}}\left[{\bf{U}}_i^{\sf{H}}{\bf{V}}_i\right]^{-1}{\bf{U}}_i^{\sf{H}}{\bf{n}}_i\\
\label{free1}
&=& {\bf{s}}_i+{1\over{H_{ii}}}{\bf{U}}_i^{\sf{H}}{\bf{n}}_i,
\end{eqnarray}
where the second equation is based on condition (\ref{TIM1}) to eliminate the interference contributed by other messages, and the last equation is obtained by setting ${\bf{U}}_i^{\sf{H}}{\bf{V}}_i={\bf{I}}$. Based on (\ref{free1}), we have the following parallel interference-free
channels for each desired symbol steam: 
\begin{eqnarray}
\label{pfree}
\tilde y_{i,m}=s_{i,m}+\tilde n_{i,m}, m\in\{1,2,\dots, M_i\},
\end{eqnarray}
where $\tilde{\bf{y}}_i=[\tilde y_{i,m}]$, ${\bf{s}}_i=[s_{i,m}]$ and ${1\over{H_{ii}}}{\bf{U}}_i^{\sf{H}}{\bf{n}}_i=[\tilde
n_{i,m}]$. As
each interference-free channel contributes $1/N$ DoF, we have $M_i/N$ DoFs
for the desired message $W_i$. Note that for the generic invertible matrix ${\bf{U}}_i^{\sf{H}}{\bf{V}}_i$, we can always obtain the parallel interference-free channels (\ref{pfree}) with different noise terms to achieve $M_i/N$ DoF in the high SNR regime.   
\end{remark}}}

Given the number of data streams $M_1, \dots, M_K$, to maximize the achievable DoFs, i.e., $M_1/N,\dots, M_K/N$, it is  equivalent to minimizing $N$, or the rank of the
matrix ${\bf{X}}$, subject to constraint (\ref{TIM}). Thus the linear TIM problem can be reformulated as the following matrix completion problem \cite{Hassibi_TIA2014,Yuanming_ISIT2015TIA}:
\begin{eqnarray}
\label{tim_lrmc}
\mathscr{P}:
\mathop {\rm{minimize}}_{{\bf{X}}\in\mathbb{R}^{M\times M}}&& {\rm{rank}}({\bf{X}})\nonumber\\
{\rm{subject~to}}&&\mathcal{P}_{\Omega}({\bf{X}})={\bf{I}}_M.
\end{eqnarray}
Note that, we only need to consider problem $\mathscr{P}$ in the real field without losing any performance in terms of achievable DoFs, as the problem parameter ${\bf{I}}_M$ is a real matrix {\rev{and the matrices ${\bf{U}}_i^{\sf{H}}{\bf{V}}_j,
\forall i\ne j, (i,j)\notin \mathcal{V}$ can be further restricted to the real field, whose corresponding signals will not contribute any interference}}. {\rev{Let ${\bf{X}}^{\star}$ be the solution of problem $\mathscr{P}$, and we can extract the precoding matrices ${\bf{V}}_j$'s
and decoding
matrices ${\bf{U}}_i$'s by performing matrix factorization as ${\bf{X}}^{\star}={\bf{U}}^{\sf{H}}{\bf{V}}=[{\bf{U}}_i^{\sf{H}}{\bf{V}}_j]$, which can be obtained by  the QR decomposition for matrix ${\bf{X}}^{\star}$
using the Gram-Schmidt process.}}

\begin{figure}[t]
  \centering
  \includegraphics[width=0.95\columnwidth]{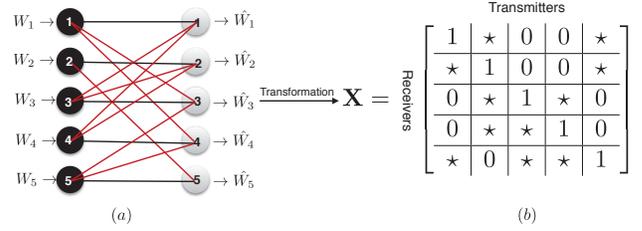}
 \caption{(a) The topological interference management problem in a partially connected
network with no CSI at transmitters (except the network topology information).
The desired channel links are black and interference links are red. (b) Associated
 incomplete matrix with ``$\star$" representing arbitrary values. For example, as there is no interference from transmitter 2 to receiver 1, $X_{12}={\bf{u}}_1^{\sf{H}}{\bf{v}}_2$ can take any value; while $X_{13}={\bf{u}}_1^{\sf{H}}{\bf{v}}_3$ must be 0 as it represents the equivalent interference channel from user 3 to user 1.}
 \label{TIM_example}
 \end{figure}

The achievable DoFs will then be given by $M_1/{\rm{rank}}({\bf{X}}^{\star}), \dots, M_1/{\rm{rank}}({\bf{X}}^{\star})$ with ${\bf{X}}^{\star}$ as the optima of problem $\mathscr{P}$. This LRMC approach for the TIM problem has been presented
in \cite{Hassibi_TIA2014,Yuanming_ISIT2015TIA} for the single data stream
transmission with the performance metric as the symmetric DoF, i.e., $M_{i}=1, \forall i$. 
While problem $\mathscr{P}$ in (\ref{tim_lrmc}) provides a clean formulation of the TIM problem, compared to existing matrix completion problems, unique challenges arise with the poorly structured affine constraint, as will be illustrated in the next subsection. An example of the idea of transforming the TIM problem to the corresponding matrix completion problem is illustrated in Fig. {\ref{TIM_example}}. For this special case, we can rewrite the  conditions (\ref{TIM1}) and (\ref{TIM2}) as the incomplete matrix ${\bf{X}}=[X_{ij}]$ with $X_{ij}={\bf{u}}_i^{\sf{H}}{\bf{v}}_j$.

\subsection{Problem Analysis}
The problem of rank minimization with affine constraints  has received enormous attention in areas such as collaborative
filtering,
statistical machine learning, as well as image and signal processing \cite{Candes_2009exactMC,Candes2013phase}. Recently, the rank minimization approach has been proposed to solve the design problem of transmit and receive beamaformers for interference alignment in MIMO interference channels \cite{Dimakis_TSP2012IArank}. 
However, the non-convex rank objective function in the LRMC problem $\mathscr{P}$ makes it NP-hard.  Enormous progress has been made recently to address the NP-hardness of the LRMC problem with elegant theoretical results using convex relaxation approaches \cite{Candes_2009exactMC} and non-convex optimization approaches \cite{Jain_2013lowAltmin}. However, most of the results highly rely on the assumptions of well structured affine constraints, e.g., the set $\Omega$ is uniformly sampled \cite{Candes_2009exactMC, Jain_2013lowAltmin} and the original matrix to be recovered is incoherent \cite{Candes_2009exactMC}. 

Unfortunately, with the poorly structured affine constraint in problem $\mathscr{P}$, none of the above standard assumptions in the literature  is satisfied. This brings unique challenges for solving and analyzing the LRMC problem $\mathscr{P}
$ for topological interference management. In this subsection, we will first review the existing algorithms for the LRMC problem and then motivate our proposed algorithm based on Riemannian optimization \cite{Absil_2009optimizationonManifolds}.       

\subsubsection{Nuclear Norm Minimization}
Let ${\bf{X}}=\sum_{i=1}^M\sigma_i{\bf{u}}_i{\bf{v}}_i^{\sf{H}}$ be the singular value decomposition (SVD)
of the matrix $\bf{X}$ with $\sigma_i$'s as the singular values and ${\bf{u}}_i$'s and ${\bf{v}}_i$'s as the left and right singular vectors, respectively. The rank function ${\rm{rank}}({\bf{X}})=\|{\bs{\sigma}}\|_{0}$ with ${\bs{\sigma}}=(\sigma_1,\dots, \sigma_M)$ is often relaxed with the nuclear norm $\|{\bf{X}}\|_{*}=\|{\bs{\sigma}}\|_{1}$ as a convex surrogate \cite{Candes_2009exactMC}, which can be regarded as an analogy with convex $\ell_1$-norm relaxation of the non-convex  $\ell_0$-norm in sparse signal recovery. If we apply this relaxation to  problem $\mathscr{P}$, it will give the following problem,
\begin{eqnarray}
\label{nunorm}
\mathop {\rm{minimize}}&& \|{\bf{X}}\|_{*}\nonumber\\
{\rm{subject~to}}&&\mathcal{P}_{\Omega}({\bf{X}})={\bf{I}}_M.
\end{eqnarray}

Unfortunately,
based on the following fact \cite{Hassibi_TIA2014}:
\begin{eqnarray}
|{\rm{Tr}}({\bf{X}})|&=&\left|{\rm{Tr}}\left(\sum\nolimits_{i}\sigma_i{\bf{u}}_i{\bf{v}}_i^{\sf{H}}\right)\right|=\left|\sum\nolimits_{i}{\rm{Tr}}\left(\sigma_i{\bf{u}}_i{\bf{v}}_i^{\sf{H}}\right)\right|\nonumber\\
&=&\left|\sum\nolimits_{i}\sigma_i{\bf{v}}_i^{\sf{H}}{\bf{u}}_i\right|\le\sum\nolimits_{i}\sigma_i|{\bf{v}}_i^{\sf{H}}{\bf{u}}_i|\nonumber\\
&\le& \sum\nolimits_{i}\sigma_i=\|{\bf{X}}\|_*,
\end{eqnarray} 
problem (\ref{nunorm}) will always return the solution ${\bf{X}}={\bf{I}}_M$, which is full rank. As a consequence, with the poorly structured affine constraint in problem $\mathscr{P}$, the nuclear norm based convex relaxation approach is inapplicable to problem $\mathscr{P}$.  

\subsubsection{Alternating Optimization Approaches}
Alternating minimization \cite{Jain_2013lowAltmin,Wotao_2012solvingLR} is another  popular non-convex optimization approach to solve the LRMC problem. Specifically, the alternating minimization approach involves expressing the unknown rank-$r$ matrix ${\bf{X}}$ as the product of two smaller matrices ${\bf{U}}{\bf{V}}^T$, where ${\bf{U}}\in\mathbb{R}^{M\times r}$ and ${\bf{V}}\in\mathbb{R}^{M\times r}$, such that the low-rank property of the matrix $\bf{X}$ is automatically satisfied. Based on this factorization, the original LRMC problem  $\mathscr{P}$ with the optimal rank as a prior information can be reformulated as the following non-convex optimization problem: 
\begin{eqnarray}
\label{altminfac}
\mathop {\rm{minimize}}_{{\bf{U}}\in\mathbb{R}^{M\times r}, {\bf{V}}\in\mathbb{R}^{M\times r}}&& \|\mathcal{P}_{\Omega}({\bf{U}}{\bf{V}}^T)-{\bf{I}}_M\|_F^2.
\end{eqnarray}
The alternating minimization algorithm for problem (\ref{altminfac}) consists of alternatively solving for ${\bf{U}}$ and ${\bf{V}}$ while fixing the other factor.  
 
 However, the fixed-rank based alternating minimization approach has a low convergence rate \cite{Vandereycken2013low,Mishra_2014r3mc}. It also fails to utilize the second-order  information to improve the convergence rate, e.g., the Hessian of the objective function. Moreover, it  requires the optimal rank as a prior information, which is, however, not available in problem $\mathscr{P}$.

\subsection{Riemannian Pursuit}
In this paper, we propose a Riemannian pursuit algorithm based on the Riemannian optimization technique \cite{Absil_2009optimizationonManifolds} to solve the LRMC problem $\mathscr{P}$ by alternatively performing the fixed-rank optimization and rank increase, thereby detecting the minimum rank of matrix $\bf{X}$ in problem $\mathscr{P}$. The proposed algorithm is described as Algorithm {\ref{rpcode}}. It will well address the limitations of the existing fixed-rank based methods \cite{Hassibi_TIA2014,Wotao_2012solvingLR,Jain_2013lowAltmin,Jain_2010guaranteedRMSVP} by 
\begin{enumerate}
\item Designing efficient algorithms for fixed-rank optimization to minimize the squared errors
of the affine constraint in problem $\mathscr{P}$;
\item Designing an effective rank increasing strategy to find good initial points in the procedure of rank pursuit, thereby detecting the minimum
rank of matrix $\bf{X}$ such that the affine constraint
in problem $\mathscr{P}$ is satisfied.
\end{enumerate}

Specifically, by fixing the rank of matrix ${\bf{X}}$ as $r~(1\le r\le M)$, we propose to solve the following smooth fixed-rank constrained  optimization problem,
\begin{eqnarray}
\mathscr{P}_{r}:
\mathop {\rm{minimize}}_{{\bf{X}}\in\mathcal{M}_{r}}&& f({\bf{X}}),
\end{eqnarray}
where $ f({\bf{X}}):={1\over{2}}\|\mathcal{P}_{\Omega}({\bf{X}})-{\bf{I}}_M\|_F^2$
is the cost function representing the squared errors of the affine constraint
in problem $\mathscr{P}$, and $\mathcal{M}_r$ is a smooth ($C^{\infty}$)
manifold given by
\begin{eqnarray}
\label{lrm}
\mathcal{M}_r:=\{{\bf{X}}\in\mathbb{R}^{M\times M}: {\rm{rank}}({\bf{X}})=r\}.
\end{eqnarray}

Observing that the least-squared cost function in problem $\mathscr{P}_{r}$ is also smooth, we thus adopt the Riemannian optimization technique \cite{Absil_2009optimizationonManifolds}
to solve it. Riemannian optimization has recently gained popularity due to its capability of exploiting the geometry of well structured search spaces based on matrix factorization \cite{Absil_2009optimizationonManifolds,Vandereycken2013low,Nicolas_2011rtrmc,Meyer2011linear,Mishra_2014fixedrank,Mishra_2014r3mc,Mishra_RP2014}, thereby being competitive with alternative approaches, e.g., convex relaxation and alternating minimization. In particular, the Riemannian optimization is the generalization
of standard unconstrained optimization, where the search space is $\mathbb{R}^n$,
to optimization of a smooth objective function on the search space of a Riemannian manifold. The details of Riemannian optimization for the fixed-rank optimization problem $\mathscr{P}_r$ will be presented in Section {\ref{Manopt}}. 

The rank increasing strategy plays an important role in the proposed algorithm. In particular, by embedding the critical point ${\bf{X}}^{[r]}$ in the current iteration into the manifold $\mathcal{M}_{r+1}$ in the next iteration, we propose an efficient  rank increasing strategy to generate good initial points and guarantee monotonic decrease of the objective function for fixed-rank optimization in the procedure of rank pursuit. This is achieved by exploiting the structures of the low-rank matrix varieties and the manifold geometry of fixed-rank matrices. The rank increasing strategy will be presented in Section {\ref{rankinc}}.

\begin{algorithm}[t]
\caption{Riemannian Pursuit (RP) for LRMC problem $\mathscr{P}$}
\begin{algorithmic}[1]
\STATE {\textbf{Input}}: $M$, $\Omega$, desired
accuracy $\epsilon$.
\STATE Initialize: ${\bf{X}}_{0}^{[1]}\in\mathbb{R}^{M\times M}, r=1$.
\WHILE{not converged}
\STATE Compute a critical point ${\bf{X}}^{[r]}$ for the smooth fixed rank-$r$
problem $\mathscr{P}_{r}$ 
 with initial point ${\bf{X}}_{0}^{[r]}$ with the Riemannian optimization
algorithm in Section {\ref{Manopt}}. 
\STATE Update the rank $r\leftarrow r+1$. Compute the initial point ${\bf{X}}_{0}^{[r]}$
for the next iteration based on the rank increasing algorithm in Section {\ref{rankinc}}.

\ENDWHILE 
\STATE {\textbf{Output}}: ${\bf{X}}^{[r]}$ and the detected minimum rank $r$.
\end{algorithmic}
\label{rpcode}
\end{algorithm}

\section{A Riemannian Optimization Framework for Smooth Fixed-Rank Optimization}
\label{Manopt}
In this section, we present a versatile framework  of Riemannian optimization for the fixed-rank matrix completion problem $\mathscr{P}_r$. It is performed on the quotient manifolds and exploits the {symmetry} structure (i.e., the quotient manifold geometry) in the search space of the fixed-rank constraint and the Hessian of the {least-squares structure} of the cost function. Specifically, the problem structures will be presented in Section {\ref{pstr}}. The framework of Riemannian optimization on the quotient manifolds will be demonstrated in Section {\ref{manopt}}. In particular, the matrix representations of all the optimization ingredients and algorithm implementation details will be provided in Section {\ref{qman}} and in Section {\ref{roalg}}, respectively.  

\subsection{Problem Structures}
\label{pstr}
To develop efficient algorithms for the smooth fixed-rank optimization problem $\mathscr{P}_r$, we exploit two fundamental structures: one is the symmetry in the fixed-rank constraint; and the other is the least-squares structure of the cost function. All the structures will be incorporated into the  Riemannian optimization framework. 
\subsubsection{Matrix Factorization and Quotient Manifold}
The set $\mathcal{M}_r$ is known to be a smooth submanifold of dimension $(2M-r)
r$ embedded in the Euclidean space $\mathbb{R}^{M\times M}$ \cite{Vandereycken2013low}.
Based on the SVD-type factorization, we represent ${\bf{X}}\in\mathcal{M}_r$
as \cite{Mishra_2014fixedrank}
\begin{eqnarray}
\label{matfac}
{\bf{X}}={\bf{U}}{\bf{\Sigma}}{\bf{V}}^{T},
\end{eqnarray}
where ${\bf{U}}, {\bf{V}}\in{\rm{St}}(r, M)$ and ${\bf{\Sigma}}\in {\rm{GL}}(r)$.
Here, ${\rm{St}}(r,M)=\{{\bf{Y}}\in\mathbb{R}^{M\times r}: {\bf{Y}}^{T}{\bf{Y}}={\bf{I}}_r\}$
denotes the \emph{Stiefel manifold} of orthonormal $M\times r$ matrices and ${\rm{GL}}(r)=\{{\bf{Y}}\in\mathbb{R}^{r\times
r}: {\rm{rank}}({\bf{Y}})=r\}$ is the set of all $r\times r$ invertible matrices.
However, the factorization in (\ref{matfac}) is not unique as we have the
symmetry structures
${\bf{X}}=({\bf{U}}{\bf{Q}}_U)({{\bf{Q}}_U^T\bf{\Sigma}}{\bf{Q}}_V)({\bf{V}}{\bf{Q}}_V)^{T},
{\bf{Q}}_U, {\bf{Q}}_V\in\mathcal{Q}(r)$,
where ${\mathcal{Q}}(r)$ is  the set of all $r\times r$ orthogonal matrices
given by $\mathcal{O}(r)=\{{\bf{Q}}\in\mathbb{R}^{r\times r}: {\bf{Q}}^{T}{\bf{Q}}={\bf{I}}_r\}$.
Therefore, the search space for problem $\mathscr{P}_r$ should be the set of
equivalence classes as follows:
\begin{eqnarray}
\label{ecla}
\!\!\![{\bf{X}}]=\{({\bf{U}}{\bf{Q}}_U, {{\bf{Q}}_U^T\bf{\Sigma}}{\bf{Q}}_V,
{\bf{V}}{\bf{Q}}_V):{\bf{Q}}_U, {\bf{Q}}_V\in\mathcal{Q}(r) \}.
\end{eqnarray}
In particular, denote the \emph{computation space}
(or the total space) as 
${\mathcal{M}}_r:={\rm{St}}(r,M)\times {\rm{GL}}(r)\times {\rm{St}}(r,M)$.
The abstract \emph{quotient} space $\mathcal{M}_r/\sim$ makes the optima isolated as 
$\mathcal{M}_r/\sim:=\mathcal{M}_r/(\mathcal{O}(r)\times
\mathcal{O}(r))$,
where $\mathcal{O}(r)\times \mathcal{O}(r)$ is the {fiber
space} and $\sim$ represents the equivalence relation. More details of
the quotient manifolds can be found in \cite{Absil_2009optimizationonManifolds}.
As the quotient manifold $\mathcal{M}_r/\sim$ is an abstract space, to design
algorithms, the matrix representation in the computation space is required.

\subsubsection{Least-Squares Structures and Riemannian Metric}
To optimize on the abstract search
space $\mathcal{M}_r/\sim$, a \emph{Riemannian metric} in the computation
space $\mathcal{M}_r$ is required
such that $\mathcal{M}_r/\sim$ is a Riemannian submersion \cite[Section 3.6.2]{Absil_2009optimizationonManifolds}. In particular, the only constraint imposed on the metric is that it should be invariant
along the set of equivalence classes $[\bf{X}]$ (\ref{ecla}). The
Riemannian metric $g_{\bf{X}}: T_{\bf{X}}\mathcal{M}_r\times T_{\bf{X}}\mathcal{M}_r\rightarrow
\mathbb{R}$ defines an inner product between the tangent vectors on the \emph{tangent
space} $T_{\bf{X}}\mathcal{M}_r$ in the computation space $\mathcal{M}_r$. 

Furthermore, by encoding the Hessian (the second-order information) of the cost function into the metric $g_{\bf{X}}$, superlinear convergence rates can be achieved for the first-order optimization algorithms \cite{Wright_2006numericalopt, Mishra_RP2014}. However, calculating the Hessian of the cost function $f$ in problem $\mathscr{P}$ is computationally costly. We thus propose a valid Riemannian metric based on the block diagonal approximation of the Hessian of the simplified cost function as presented in the following proposition.

\begin{proposition}[Riemannian Metric]
\label{proprm} 
By exploiting the second order information of the least-squares cost function, the Riemannian metric $g_{\bf{X}}: T_{\bf{X}}\mathcal{M}_r\times T_{\bf{X}}\mathcal{M}_r\rightarrow
\mathbb{R}$ is given by
\begin{eqnarray}
\label{rmetric}
g_{\bf{X}}({\bs{\xi}}_{\bf{X}}, {\bs{\zeta}}_{\bf{X}})&=&\langle {\bs{\xi}}_{{U}},
{\bs{\zeta}}_{{U}}{\bf{\Sigma}}{\bf{\Sigma}}^T\rangle+\langle
{\bs{\xi}}_{{\Sigma}}, {\bs{\zeta}}_{{\Sigma}}\rangle+\nonumber\\
&&\langle {\bs{\xi}}_{{V}},
{\bs{\zeta}}_{{V}}{\bf{\Sigma}}^T{\bf{\Sigma}}\rangle,
\end{eqnarray}
where ${\bs{\xi}}_{\bf{X}}:=({\bs{\xi}}_{{U}}, {\bs{\xi}}_{{\Sigma}},
{\bs{\xi}}_{{V}})\in T_{\bf{X}}\mathcal{M}_r, {\bs{\zeta}}_{\bf{X}}:=({\bs{\zeta}}_{{U}},
{\bs{\zeta}}_{{\Sigma}},
{\bs{\zeta}}_{{V}})\in T_{\bf{X}}\mathcal{M}_r$ and ${\bf{X}}:=({\bf{U}},
{\bf{\Sigma}}, {\bf{V}})$. 
\begin{IEEEproof}
Please refer to Appendix {\ref{apprm}} for details. 
\end{IEEEproof}
\end{proposition}
Note that, different from the conventional metric \cite{Meyer2011linear},
which only
takes the search space into consideration, the novel metric (\ref{rmetric}) can
encode  the second-order information of the cost
function, thus leads to a faster convergence speed for the first-order algorithms \cite{Mishra_RP2014, Wright_2006numericalopt}. This will
be further justified in the simulation section.

\subsection{Riemannian Optimization on Quotient Manifolds}
\label{manopt} 
The main idea of Riemannian optimization is to encode the constraints on the manifold into the search space, and then perform descent on this manifold search space rather than  in the ambient Euclidean space. {\rev{In particular, the Euclidean gradient and Euclidean Hessian need to
be converted to the Riemannian gradient and Riemannian Hessian, respectively,
to implement the conjugate gradient method and trust-region method in the
Riemannian optimization framework. This will be explicitly presented in Section {\ref{qman}}.}} For the quotient manifold $\mathcal{M}_r/\sim$, the abstract geometric objects call for concrete matrix representations in the computation space $\mathcal{M}_r$, which is achieved by
the principle of the \emph{Riemannian
submersion} \cite[Section 3.6.2]{Absil_2009optimizationonManifolds}.
Therefore, essentially, the algorithms are implemented in the computation space.  Specifically, with the Riemannian
metric (\ref{rmetric}), the quotient manifold $\mathcal{M}_r/\sim$ is \emph{submersed}
into $\mathcal{M}_r$. We now have the \emph{Riemannian quotient manifold} as follows:
{\rev{\begin{definition}[Riemannian Quotient Manifold {\cite[Section 3.6.2]{Absil_2009optimizationonManifolds}}]
Endowed with the Riemannian metric (\ref{rmetric}),
$\mathcal{M}_r/\sim$ is called a \emph{Riemannian quotient manifold} of $\mathcal{M}_r$.
\end{definition}}}

Let $T_{[\bf{X}]}(\mathcal{M}_r/\sim)$ denote the abstract tangent space
in the quotient manifold ${\mathcal{M}}_r/\sim$, which has the matrix representation
in $T_{\bf{X}}\mathcal{M}_r$. The abstract tangent vectors in $T_{[\bf{X}]}(\mathcal{M}_r/\sim)$
are restricted to the directions
that do not produce a displacement
along the equivalence class $[{\bf{X}}]$ (\ref{ecla}). 
This is achieved by decomposing the tangent space $T_{{\bf{X}}}{\mathcal{M}_r}$
in the computation space 
into complementary spaces as follows:
$T_{{\bf{X}}}{\mathcal{M}}_r=\mathcal{V}_{{\bf{X}}}{\mathcal{M}}_r\otimes\mathcal{H}_{{\bf{X}}}{\mathcal{M}}_r$,
where $\mathcal{V}_{\bf{X}}\mathcal{M}_r$ and $\mathcal{H}_{\bf{X}}\mathcal{M}_r$
are the \emph{vertical space} and \emph{horizontal space}, respectively.
In particular, the {horizontal space} $\mathcal{H}_{\bf{X}}\mathcal{M}_r$,
which is the orthogonal complement of $\mathcal{V}_{\bf{X}}\mathcal{M}_r$ in
the sense {of the Riemannian metric} $g_{\bf{X}}$, provides a valid matrix representation
of the abstract tangent space $T_{[\bf{X}]}(\mathcal{M}_r/\sim)$ \cite[Section
3.5.8]{Absil_2009optimizationonManifolds}. The vertical space $\mathcal{V}_{\bf{X}}\mathcal{M}_r$
is obtained from the tangent space of the equivalence class $[{\bf{X}}]$ (\ref{ecla}).
We call it the \emph{horizontal lift} given that any element in the
abstract tangent space ${\bs{\xi}}_{[\bf{X}]}\in T_{[\bf{X}]}(\mathcal{M}_r/\sim)$
has a unique element in the horizontal space ${\bs{\xi}}_{\bf{X}}\in\mathcal{H}_{\bf{X}}\mathcal{M}_r$.

As $g_{\bf{X}}$ is constrained to be invariant along the equivalent class $[\bf{X}]$ (\ref{ecla}), it can define a Riemannian metric $g_{[\bf{X}]}({\bs{\xi}}_{[{\bf{X}}]},
{\bs{\zeta}}_{[{\bf{X}}]}): T_{[\bf{X}]}(\mathcal{M}_r/\sim)\times T_{[\bf{X}]}(\mathcal{M}_r/\sim)\rightarrow
\mathbb{R}$ in the quotient space $\mathcal{M}_r/\sim$ as 
$g_{[\bf{X}]}({\bs{\xi}}_{[{\bf{X}}]}, {\bs{\zeta}}_{[{\bf{X}}]}):=g_{\bf{X}}({\bs{\xi}}_{\bf{X}},
{\bs{\zeta}}_{\bf{X}})$,
where ${\bs{\xi}}_{[{\bf{X}}]}, {\bs{\zeta}}_{[{\bf{X}}]}\in T_{[\bf{X}]}(\mathcal{M}_r/\sim)$
 and ${\bs{\xi}}_{\bf{X}}, {\bs{\zeta}}_{\bf{X}}\in\mathcal{H}_{\bf{X}}\mathcal{M}_r$
are the horizontal lifts or matrix representations of ${\bs{\xi}}_{[{\bf{X}}]}$
and ${\bs{\zeta}}_{[{\bf{X}}]}$. Note that both ${\bs{\xi}}_{\bf{X}}$ and ${\bs{\zeta}}_{\bf{X}}$
 belong to the tangent space $T_{\bf{X}}\mathcal{M}_r$. In summary, we have \emph{Riemannian submersion} as follows:
{\rev{
\begin{definition}[Riemannian Submersion {\cite[Section 3.6.2]{Absil_2009optimizationonManifolds}}]
The choice of the metric (\ref{rmetric}), which is invariant along the equivalent
class $[\bf{X}]$, and of the horizontal space $\mathcal{H}_{\bf{X}}\mathcal{M}_r$ as the orthogonal complement of $\mathcal{V}_{\bf{X}}$, in the sense of the Riemannian metric (\ref{rmetric}), makes the search space $\mathcal{M}_r/\sim$ a \emph{Riemannian submersion}.    
\end{definition}}}

\begin{figure}[t]
  \centering
  \includegraphics[width=0.95\columnwidth]{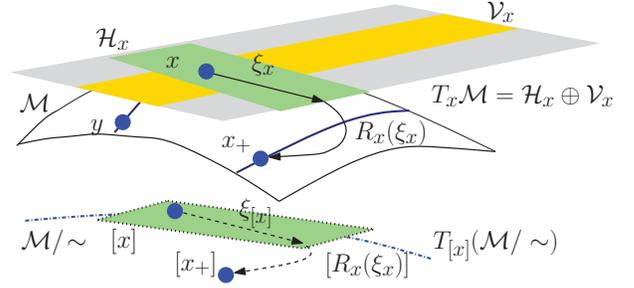}
 \caption{A schematic view of Riemannian optimization framework: abstract
geometric objects (shown in dotted line) on a quotient manifold $\mathcal{M}_r/\sim$
call for matrix representatives (shown in solid lines) in the computation
space (or total space) $\mathcal{M}_r$. The points ${\bf{x}}$ and ${\bf{y}}$ in $\mathcal{M}_r$ belong to the  same equivalence class (shown in solid blue color) and they represent  a single point $[{\bf{x}}]=\{{\bf{y}}\in\mathcal{M}_r: {\bf{y}}\sim {\bf{x}}\}$ on the quotient manifold $\mathcal{M}_r/\sim$. Figure courtesy of Mishra {\emph{et al.}} \cite{Mishra_RP2014}.}
 \label{roptfig}
 \end{figure}

Therefore, with the metric (\ref{rmetric}), the Riemannian optimization algorithms on the quotient manifold $\mathcal{M}_r/\sim$ call for matrix representation (horizontal lifts) in the computation space $\mathcal{M}_r$. Specifically, let ${\bf{\Xi}}_i\in\mathcal{H}_{{\bf{X}}_i}\mathcal{M}_r$ be the search direction at the $i$-th iteration. Define   $\mathcal{R}_{\bf{X}}: \mathcal{H}_{\bf{X}}\mathcal{M}_r\rightarrow
\mathcal{M}_r$
as the \emph{retraction} mapping operator that maps the element in the
horizontal space ${\bs{\Xi}}_{i}\in\mathcal{H}_{\bf{X}}\mathcal{M}_r$
to the points on the computation space $\mathcal{M}_r$. The Riemannian optimization framework for the smooth optimization problem $\mathscr{P}_r$ is presented in Algorithm {\ref{manoptcode}} and the corresponding schematic view  is shown in Fig. {\ref{roptfig}}. {\rev{In particular, the parameter $\alpha_i$ in Algorithm {\ref{manoptcode}} denotes the step size, which we will explain in Section {\ref{roalg}}.}} 
\begin{algorithm}
\caption{A Riemannian Optimization Framework for the Fixed-Rank  Optimization Problem
$\mathscr{P}_{r}$}
\begin{algorithmic}[1]
\STATE {\textbf{Input}}: $M$,  $r$, $\Omega$, desired
accuracy $\varepsilon$.
\STATE Initialize: ${\bf{X}}_0={\bf{X}}^{\textrm{initial}}, {\bf{\Xi}}_0={\bf{0}},
i=0$.
\WHILE{not converged}
\STATE Compute the search direction ${\bf{\Xi}}_i\in \mathcal{H}_{{\bf{X}}_i}\mathcal{M}_{
r}$.
\STATE Update ${\bf{X}}_{i+1}={\mathcal{R}}_{{{\bf{X}}_{i}}}(\alpha_i{\bf{\Xi}}_i)$.
Update $i=i+1$. 
\ENDWHILE 
\STATE {\textbf{Output}}: ${\bf{X}}^{\star}={\bf{X}}_{i}$.
\end{algorithmic}
\label{manoptcode}
\end{algorithm} 

\subsection{ Quotient Manifold Representation}
\label{qman}
In this subsection, we derive  the concrete matrix representations (horizontal lifts) in the computation space $\mathcal{M}_r$ for abstract geometric objects on the quotient manifold $\mathcal{M}_r/\sim$, thereby implementing the Riemannian optimization algorithms.  
\subsubsection{Riemannian Gradient}
To design an algorithm using the conjugate gradient method on he quotient space $\mathcal{M}_r/\sim$, we need to define the \emph{Riemannian gradient} ${\rm{grad}}_{[\bf{X}]}f$ for the objective function $f(\bf{X})$ on this space, which is the generalization of the Euclidean gradient $\nabla f({\bf{X}})=\mathcal{P}_{\Omega}({\bf{X}})-{\bf{I}}_M$ of $f(\bf{X})$.  To achieve this goal, we first provide the following proposition on the matrix representation of the abstract tangent space $T_{[\bf{X}]}(\mathcal{M}_r/\sim)$.

\begin{proposition}[Horizontal Space]
\label{prohs}
The \emph{horizontal space} $\mathcal{H}_{\bf{X}}\mathcal{M}_r$,
which is any complementary subspace of $\mathcal{V}_{\bf{X}}\mathcal{M}_r$
in the sense of the {Riemannian metric} $g_{\bf{X}}$ (\ref{rmetric}), provides
a valid matrix representation of the abstract tangent space $T_{[\bf{X}]}(\mathcal{M}_r/\sim)$
as $\mathcal{H}_{\bf{X}}\mathcal{M}_r\!=\!\{{\bs{\eta}}_{\bf{X}}\in
T_{\bf{X}}\mathcal{M}_r: {\bf{S}}_1~{\rm{and}}~{\bf{S}}_2~{\rm{are
~symmetric}}\}$, 
where ${\bf{S}}_1={\bf{\Sigma}}{\bf{\Sigma}}^{T}{\bs{\eta}}_{\bf{U}}^{T}{\bf{U}}-{\bf{\Sigma}}{\bs{\eta}}_{\bf{\Sigma}}^T$
and ${\bf{S}}_2={\bf{\Sigma}}^{T}{\bf{\Sigma}}{\bs{\eta}}_{\bf{V}}^T{\bf{V}}+{\bs{\eta}}_{\bf{\Sigma}}^T{\bf{\Sigma}}$.
\begin{IEEEproof}
Please refer to Appendix {\ref{apphs}} for details. 
\end{IEEEproof}
\end{proposition} 

To compute the Riemannian gradient, we need to  define two projection operators: {tangent space projection} and {horizontal space projection}. Specifically, the tangent space projection is the operator that  projects the ambient space onto the tangent space.    
\begin{proposition}[Tangent Space Projection]
\label{protp}
The tangent space projection operator $P_{T_{\bf{X}}\mathcal{M}_r}: \mathbb{R}^{M\times
r}\times \mathbb{R}^{r\times r}\times \mathbb{R}^{M\times r}\rightarrow T_{\bf{X}}\mathcal{M}_r$ that projects the ambient space $\mathbb{R}^{M\times
r}\times \mathbb{R}^{r\times r}\times \mathbb{R}^{M\times r}$ onto the tangent
space $T_{\bf{X}}\mathcal{M}_r$ is given by:
\begin{eqnarray}
\label{tanpro}
P_{T_{\bf{X}}\mathcal{M}_r}({\bf{A}}_{{U}}, {\bf{A}}_{{\Sigma}}, {\bf{A}}_{{V}})=({\bs{\xi}}_{{U}},
{\bs{\xi}}_{{\Sigma}}, {\bs{\xi}}_{{V}}),
\end{eqnarray}
where 
${\bs{\xi}}_{{U}}={\bf{A}}_{{U}}-{\bf{U}}{\bf{B}}_{{U}}({\bf{\Sigma}}{\bf{\Sigma}}^T)^{-1}$,
${\bs{\xi}}_{{\Sigma}}= {\bf{A}}_{{U}}$,
${\bs{\xi}}_{{V}}= {\bf{A}}_{{V}}-{\bf{V}}{\bf{B}}_{{V}}({\bf{\Sigma}}^T{\bf{\Sigma}})^{-1}$.
Here, ${\bf{B}}_{{U}}$ and ${\bf{B}}_{{V}}$ are symmetric matrices
of size $r\times r$ that are obtained by solving the Lyapunov equations
\begin{eqnarray}
\!\!\!\!\!\!{\bf{\Sigma}}{\bf{\Sigma}}^{T}{\bf{B}}_{{U}}+{\bf{B}}_{{U}}{\bf{\Sigma}}{\bf{\Sigma}}^{T}&=&{\bf{\Sigma}}{\bf{\Sigma}}^{T}({\bf{U}}^T{\bf{A}}_{{U}}+{\bf{A}}_{{U}}^{T}{\bf{U}}){\bf{\Sigma}}{\bf{\Sigma}}^{T},\\
\!\!\!\!\!\!{\bf{\Sigma}}^T{\bf{\Sigma}}{\bf{B}}_{{V}}+{\bf{B}}_{{V}}{\bf{\Sigma}}^T{\bf{\Sigma}}&=&{\bf{\Sigma}}^T{\bf{\Sigma}}({\bf{V}}^T{\bf{A}}_{{V}}+{\bf{A}}_{{V}}^{T}{\bf{V}}){\bf{\Sigma}}^T{\bf{\Sigma}}.
\label{lyapeq1}
\end{eqnarray} 
\begin{IEEEproof}
Please refer to Appendix {\ref{apptp}} for details.
\end{IEEEproof}
\end{proposition}

The horizontal space projection is the operator that  extracts the horizontal component of the
tangent vector by projecting the tangent space onto the horizontal space.
\begin{proposition}[Horizontal Space Projection] 
\label{prohp}
The horizontal space projection operator $\Pi_{\mathcal{H}_{{\bf{X}}}{\mathcal{M}}_r}:T_{\bf{X}}\mathcal{M}_r\rightarrow \mathcal{H}_{\bf{X}}\mathcal{M}_r$ that projects the tangent space $T_{\bf{X}}\mathcal{M}_r$ onto the horizontal space $\mathcal{H}_{\bf{X}}\mathcal{M}_r$ is given by
$\Pi_{\mathcal{H}_{{\bf{X}}}{\mathcal{M}}_r}({\bs{\xi}}_{\bf{X}})=({\bs{\zeta}}_{{U}},
{\bs{\zeta}}_{{\Sigma}},{\bs{\zeta}}_{{V}})$,
where 
${\bs{\zeta}}_{{U}}={\bs{\xi}}_{{U}}-{\bf{U}}{\bf{\Theta}}_1$,
${\bs{\zeta}}_{{\Sigma}}={\bs{\xi}}_{{\Sigma}}+{\bf{\Theta}}_1{\bf{\Sigma}}-{\bf{\Sigma}}{\bf{\Theta}}_2$,
${\bs{\zeta}}_{{V}}={\bs{\xi}}_{{V}}-{\bf{V}}{\bf{\Theta}}_2$.
Here, ${\bs{\Theta}}_1$ and ${\bs{\Theta}}_2$ are skew-symmetric matrices of size
$r\times r$ that are obtained by solving the coupled system of Lyapunov equations
\begin{eqnarray}
\label{coupledly1}
{\bf{\Sigma}}{\bf{\Sigma}}^{T}{\bs{\Theta}}_1+{\bs{\Theta}}_1{\bs{\Sigma}}{\bf{\Sigma}}^T-{\bf{\Sigma}}{\bf{\Theta}}_2{\bf{\Sigma}}^T&=&{\rm{Skew}}({\bf{U}}^{T}{\bs{\xi}}_{{U}}{\bf{\Sigma}}{\bf{\Sigma}}^{T})+\nonumber\\
&&{\rm{Skew}}({\bf{\Sigma}}{\bs{\xi}}_{{\Sigma}}^{T}),\\
{\bf{\Sigma}}^T{\bf{\Sigma}}{\bs{\Theta}}_2+{\bs{\Theta}}_2{\bs{\Sigma}}^T{\bf{\Sigma}}-{\bf{\Sigma}}^T{\bf{\Theta}}_1{\bf{\Sigma}}&=&{\rm{Skew}}({\bf{V}}^{T}{\bs{\xi}}_{{V}}{\bf{\Sigma}}^T{\bf{\Sigma}})+\nonumber\\
&&{\rm{Skew}}({\bf{\Sigma}}^T{\bs{\xi}}_{{\Sigma}}),
\label{coupledly2}
\end{eqnarray}
where ${\rm{Skew}}(\cdot)$ extracts the skew-symmetric part of a square matrix,
i.e., ${\rm{Skew}}({\bf{C}})=({\bf{C}}-{\bf{C}}^{T})/2$.
\begin{IEEEproof}
Please refer to Appendix {\ref{apphp}} for details. 
\end{IEEEproof}
\end{proposition} 

Based on Propositions {\ref{protp}} and {\ref{prohp}}, we have the matrix representation (horizontal lift) ${\rm{grad}}_{\bf{X}}f$ of the Riemannian gradient ${\rm{grad}_{[\bf{X}]}} f$ on the quotient manifold $\mathcal{M}_r/\sim$ at ${\bf{X}}=({\bf{U}}, {\bf{\Sigma}}, {\bf{V}})$ as follows:
\begin{eqnarray}
\label{gradrm}
{\rm{grad}}_{\bf{X}}f=({\bs{\xi}}_{{U}}, {\bs{\xi}}_{{\Sigma}},
{\bs{\xi}}_{{V}}),
\end{eqnarray}
where
${\bs{\xi}}_{{U}}={\bf{A}}{\bf{V}}{\bf{\Sigma}}^{T}({\bf{\Sigma}}{\bf{\Sigma}}^{T})^{-1}-{\bf{U}}{\bf{B}}_{{U}}({\bf{\Sigma}}{\bf{\Sigma}}^{T})^{-1}$,
${\bs{\xi}}_{{\Sigma}}= {\bf{U}}^{T}{\bf{S}}{\bf{V}}$,
${\bs{\xi}}_{{V}}={\bf{A}}^T{\bf{U}}{\bf{\Sigma}}({\bf{\Sigma}}^T{\bf{\Sigma}})^{-1}-{\bf{V}}{\bf{B}}_{{V}}({\bf{\Sigma}}^T{\bf{\Sigma}})^{-1}$,
with ${\bf{A}}=\nabla f({\bf{X}})=\mathcal{P}_{\Omega}({\bf{X}})-{\bf{I}}_M$. Here, ${\bf{B}}_{{U}}$ and ${\bf{B}}_{V}$ are the solutions to the Lyapunov
equations 
\begin{eqnarray}
\label{blyp1}
{\bf{\Sigma}}{\bf{\Sigma}}^{T}{\bf{B}}_{{U}}+{\bf{B}}_{{U}}{\bf{\Sigma}}{\bf{\Sigma}}^{T}&=&2{\rm{Sym}}({\bf{\Sigma}}{\bf{\Sigma}}^{T}{\bf{U}}^T{\bf{A}}{\bf{V}}{\bf{\Sigma}}),\\
{\bf{\Sigma}}^T{\bf{\Sigma}}{\bf{B}}_{{V}}+{\bf{B}}_{{V}}{\bf{\Sigma}}^T{\bf{\Sigma}}&=&2{\rm{Sym}}({\bf{\Sigma}}^T{\bf{\Sigma}}{\bf{V}}^T{\bf{S}}^T{\bf{U}}{\bf{\Sigma}}),
\label{blyp2}
\end{eqnarray} 
where ${\rm{Sym}}(\cdot)$ extracts the symmetric part of a square matrix,
i.e., ${\rm{Sym}}({\bf{C}})=({\bf{C}}+{\bf{C}}^{T})/2$.  Please refer to Appendix {\ref{gradrmapp}} for the details on the derivation of the Riemannian gradient (\ref{gradrm}).        

\subsubsection{Riemannian Hessian}
To design second-order algorithms (e.g., the trust-region scheme) on the quotient
space $\mathcal{M}_r/\sim$, we need to define the \emph{Riemannian connection} on this space, which is the generalization of directional derivative of a
vector field on the manifold.  Let $\nabla_{{\bs{\eta}}_{\bf{X}}} {{\bs{\xi}}}_{\bf{X}}$ be the directional
derivative of the vector field ${\bs{\xi}}_{\bf{X}}\in T_{\bf{X}}\mathcal{M}_{r}$
applied in the direction ${\bs{\eta}}_{\bf{X}}\in T_{\bf{X}}\mathcal{M}_{r}$ on the computation space $\mathcal{M}_r$. Then
the matrix representation (horizontal lift) of the Riemannian connection $\nabla_{{\bs{\eta}}_{[\bf{X}]}}
{{\bs{\xi}}}_{[\bf{X}]}$ on the quotient space $\mathcal{M}_r/\sim$ with ${\bs{\eta}}_{[\bf{X}]}, {\bs{\xi}}_{[\bf{X}]}\in T_{[\bf{X}]}(\mathcal{M}_r/\sim)$ is given by $\Pi_{\mathcal{H}_{{\bf{X}}}{\mathcal{M}}_r}(\nabla_{{\bs{\eta}}_{\bf{X}}}
{{\bs{\xi}}}_{\bf{X}})$, which is the horizontal projection of the Riemannian connection onto the horizontal space.  By the \emph{Koszul} formula \cite[Theorem 5.3.1]{Absil_2009optimizationonManifolds},
the Riemannian connection is given by
\begin{eqnarray}
\label{rmcon}
\nabla_{{\bs{\eta}}_{\bf{X}}} {{\bs{\xi}}}_{\bf{X}}&=& {D}{\bs{\xi}}_{\bf{X}}[{\bs{\eta}}_{\bf{X}}]+({\bs{\theta}}_{{U}},
 {\bs{\theta}}_{{\Sigma}}, {\bs{\theta}}_{{V}}),
\end{eqnarray}
where ${D}{\bs{\xi}}_{\bf{X}}[{\bs{\eta}}_{\bf{X}}]$ is the classical Euclidean
directional derivative and
${\bs{\theta}}_{{U}}= {\bs{\eta}}_{{{U}}}{\bf{B}}_{{U}}+{\bf{U}}{\bf{B}}_{{U}}+2{\bs{\xi}}_{{U}}{\rm{Sym}}({\bs{\eta}}_{{\Sigma}}{\bf{\Sigma}}^{T})({\bf{\Sigma}}{\bf{\Sigma}}^{T})^{-1}$,
${\bs{\theta}}_{{\Sigma}}= {\bf{0}}$,
${\bs{\theta}}_{{V}}= {\bs{\eta}}_{{{V}}}{\bf{B}}_{{V}}+{\bf{V}}{\bf{B}}_{{V}}+2{\bs{\xi}}_{{V}}{\rm{Sym}}({\bs{\eta}}_{{\Sigma}}^{T}{\bf{\Sigma}})({\bf{\Sigma}}^{T}{\bf{\Sigma}})^{-1}$.
Here, ${\bf{B}}_{U}$ and ${\bf{B}}_V$ are the solutions to the Lyapunov equations (\ref{blyp1}) and (\ref{blyp2}).

Therefore, the matrix representation (horizontal lift) of the Riemannian Hessian ${\rm{Hess}}_{[\bf{X}]}f [{\bs{\xi}}_{\bf{X}}]$ on the quotient manifold $\mathcal{M}_r/\sim$ is given by
\begin{eqnarray}
\label{hessian}
{{\rm{Hess}}_{\bf{X}} f[{\bs{\xi}}_{\bf{X}}]}=\Pi_{\mathcal{H}_{{\bf{X}}}{\mathcal{M}}_r}(\nabla_{{\bs{\xi}}_{\bf{X}}}{\rm{grad}}_{\bf{X}}f),
\end{eqnarray}
where ${\rm{grad}}_{\bf{X}} f$ (\ref{gradrm}) is the Riemannian gradient in the computation space
$\mathcal{M}_r$ and the Riemannian connection is given in (\ref{rmcon}). 
\subsection{Riemannian Optimization Algorithms}
\label{roalg}
Based on the above matrix representations or horizontal lifts of the geometric objects on abstract search space $\mathcal{M}_r/\sim$, it is ready to implement the algorithms in the computation space $\mathcal{M}_r$. To trade off the convergence rate and the computational complexity, we present a first-order algorithm (i.e., the conjugate gradient method) and a second-order method (i.e., the trust-region method) in Section {\ref{cgm}} and Section {\ref{trm}}, respectively.  
\subsubsection{Conjugate Gradient Method}
\label{cgm}
In the conjugate gradient scheme, the search direction at iteration $i$ is given by
${\bf{\Xi}}_i:=-{\rm{grad}}_{{\bf{X}}_i} f+\beta_i \mathcal{T}_{{\bf{X}}_{i-1}\rightarrow{\bf{X}}_i}({\bf{\Xi}}_{i-1})$,
where ${\rm{grad}}_{{\bf{X}}_i} f\in\mathcal{H}_{\bf{X}}\mathcal{M}_r$ is the {Riemannian gradient} at point ${\bf{X}}_i\in\mathcal{M}_r$ and $\mathcal{T}_{{\bf{X}}_{i-1}\rightarrow {\bf{X}}_i}({\bs{\xi}}_{\bf{X}}):
\mathcal{H}_{{\bf{X}}_i}\mathcal{M}_r\rightarrow \mathcal{H}_{{\bf{X}}_i}\mathcal{M}_r$ is the matrix representation (the horizontal lift) of the \emph{vector transport} $\mathcal{T}_{[{\bf{X}}_{i-1}]\rightarrow
[{\bf{X}}_i]}({\bs{\xi}}_{[\bf{X}]})$  that maps tangent vectors from one tangent space $T_{[{\bf{X}}_{i-1}]}(\mathcal{M}_r/\sim)$ to another tangent space  $T_{[{\bf{X}}_{i}]}(\mathcal{M}_r/\sim)$ given by
$\mathcal{T}_{{\bf{X}}_{i-1}\rightarrow{\bf{X}}_i}({\bf{\Xi}}_{i-1})=\Pi_{\mathcal{H}_{{\bf{X}}_i}{\mathcal{M}}_r}(P_{T_{{\bf{X}}_i}\mathcal{M}_r}({\bf{\Xi}}_{i-1}))$.

Therefore, the sequence of the  iterates is given by
\begin{eqnarray}
{\bf{X}}_{i+1}&=&\mathcal{R}_{{\bf{X}}_i}(\alpha_{i}{\bs{\Xi}}_{i}),
\end{eqnarray}
where $\alpha_i$ denotes the step size satisfying the strong Wolf conditions \cite{Vandereycken_ICML2014RiemanMatrixRec,Absil_2009optimizationonManifolds} and  $\mathcal{R}_{\bf{X}}: \mathcal{H}_{\bf{X}}\mathcal{M}_r\rightarrow \mathcal{M}_r$
is the \emph{retraction} mapping operator that maps the element in the
horizontal space ${\bs{\Xi}}_{i}\in\mathcal{H}_{\bf{X}}\mathcal{M}_r$
to the points on the computation space $\mathcal{M}_r$. The product nature of
the computation space $\mathcal{M}_r$ allows to choose a retraction by simply combining
the retractions on the individual manifolds \cite[Example 4.1.3]{Absil_2009optimizationonManifolds},
$\mathcal{R}_{\bf{X}}({\bs{\xi}}_{\bf{X}})=({\rm{uf}}({\bf{U}}+{\bs{\xi}}_{{U}}),
{\bf{\Sigma}}+{\bs{\xi}}_{{\Sigma}}, {\rm{uf}}({\bf{V}}+{\bs{\xi}}_{{V}}))$,
where ${\bs{\xi}}_{\bf{X}}:=({\bs{\xi}}_U, {\bs{\xi}}_{\Sigma}, {\bs{\xi}}_V)\in\mathcal{H}_{\bf{X}}\mathcal{M}_r$ and ${\rm{uf}}(\cdot)$
extracts the orthogonal factor of a full column-rank matrix, i.e., ${\rm{uf}}({\bf{A}})={\bf{A}}({\bf{A}}^{T}{\bf{A}})^{-1/2}$.

\begin{figure}[t]
  \centering
  \includegraphics[width=0.95\columnwidth]{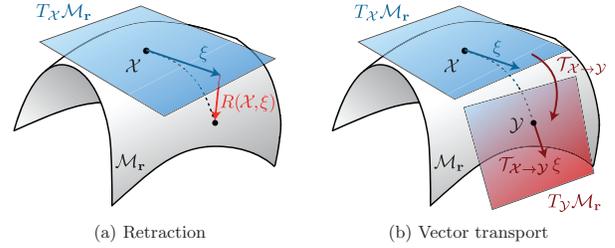}
 \caption{Visual representation of the concept of retraction and vector transport within the framework of
Riemannian optimization techniques. Figure
courtesy of Kressner {\emph{et al.}} \cite{Kressner_SV_2013}.}
 \label{vtr}
 \end{figure}

The concepts of vector transport and retraction in the total space $\mathcal{M}_r$ are illustrated on the right and left sides of Fig. {\ref{vtr}}, respectively. 

\subsubsection{Trust Region Method}
\begin{table*}[ht]
\renewcommand{\arraystretch}{1.3}
\caption{Optimization-related Ingredients for problem $\mathscr{P}_r$}
\label{opting}
\centering
\begin{tabular}{l|c}
 & $\mathscr{P}_{r}:
\mathop {\rm{minimize}}_{{\bf{X}}\in\mathcal{M}_{r}}f({\bf{X}})$\\
\hline
Matrix representation of an element ${\bf{X}}\in\mathcal{M}_r$ & ${\bf{X}}=({\bf{U}},
{\bs{\Sigma}}, {\bf{V}})$
\\
Computational space $\mathcal{M}_r$ & ${\rm{St}}(r,M)\times{\rm{GL}}(r)\times{\rm{St}}(r,
M)$\\
Quotient space & ${\rm{St}}(r,M)\times{\rm{GL}}(r)\times{\rm{St}}(r,
M)/(\mathcal{O}(r)\times\mathcal{O}(r))$\\
Metric $g_{\bf{X}}({\bs{\xi}}_{\bf{X}}, {\bs{\zeta}}_{\bf{X}})$ for ${\bs{\xi}}_{\bf{X}},
{\bs{\zeta}}_{\bf{X}}\in T_{\bf{X}}\mathcal{M}_r$ & $g_{\bf{X}}({\bs{\xi}}_{\bf{X}},
{\bs{\zeta}}_{\bf{X}})=\langle {\bs{\xi}}_{{U}},
{\bs{\zeta}}_{{U}}{\bf{\Sigma}}{\bf{\Sigma}}^T\rangle+\langle
{\bs{\xi}}_{{\Sigma}}, {\bs{\zeta}}_{{\Sigma}}\rangle+\langle {\bs{\xi}}_{{V}},
{\bs{\zeta}}_{{V}}{\bf{\Sigma}}^T{\bf{\Sigma}}\rangle$\\
Riemannian gradient ${\rm{grad}}_{\bf{X}}f$ & ${\rm{grad}}_{\bf{X}} f=({\bs{\xi}}_{U},
{\bs{\xi}}_{\Sigma}, {\bs{\xi}}_{V})$ (\ref{gradrm})\\
Riemannian Hessian ${\rm{Hess}}_{\bf{X}} f[{\bs{\xi}}_{\bf{X}}]$ & ${\rm{Hess}}_{\bf{X}}
f[{\bs{\xi}}_{\bf{X}}]=\Pi_{\mathcal{H}_{\bf{X}}\mathcal{M}_r}(\nabla_{\bs{\xi}_{\bf{X}}}{\rm{grad}}_{\bf{X}}f)$
(\ref{hessian})\\
Retraction $\mathcal{R}_{\bf{X}}({\bs{\xi}}_{\bf{X}}): \mathcal{H}_{\bf{X}}\mathcal{M}_r\rightarrow
\mathcal{M}_r$ & $({\rm{uf}}({\bf{U}}+{\bs{\xi}}_{\bf{X}}), {\bs{\Sigma}}+{\bs{\xi}}_{\Sigma},
{\rm{uf}}({\bf{V}}+{\bs{\xi}}_V))$
\end{tabular}
\end{table*}
\label{trm}
To provide quadratic convergence rate, we implement the second-order optimization algorithm based on the trust-region method \cite{Absil_2007trustregion}. In particular, in the quotient manifold $\mathcal{M}_r/\sim$, the trust-region subproblem is horizontally lifted to $\mathcal{H}_{\bf{X}}\mathcal{M}_r$ and  formulated as
\begin{eqnarray}
\label{trsd}
\mathop {\rm{minimize}}_{{\bs{\xi}}_{\bf{X}}\in \mathcal{H}_{\bf{X}}\mathcal{M}_r}&& m({\bs{\xi}}_{\bf{X}})\nonumber\\
{\rm{subject~to}}&&g_{\bf{X}}({\bs{\xi}}_{\bf{X}}, {\bs{\xi}}_{\bf{X}})\le  \delta^2,
\end{eqnarray} 
where $\delta$ is the trust-region radius and the cost function is given by
\begin{eqnarray}
m({\bs{\xi}}_{\bf{X}})&=&f({\bf{X}})+g_{\bf{X}}({\bs{\xi}}_{\bf{X}},{\rm{grad}}_{\bf{X}}f)+\nonumber\\
&&{1\over{2}} g_{\bf{X}}({\bs{\xi}}_{\bf{X}}, {\rm{Hess}}_{{\bf{X}}} f[{\bs{\xi}}_{\bf{X}}]),
\end{eqnarray}
where ${{\rm{grad}}_{\bf{X}} f}$ (\ref{gradrm}) and ${{\rm{Hess}}_{\bf{X}} f}$ (\ref{hessian}) are the horizontal lift (matrix representation) of the Riemannian gradient and Riemannian Hessian on the quotient manifold $\mathcal{M}_r/\sim$. Given the matrix representation of the search direction (\ref{trsd}), the details of the implementation of the trust-region algorithm can be found in \cite{manopt}.

In summary,  the optimization-related ingredients for problem $\mathscr{P}_r$ are provided  in Table {\ref{opting}}.

\section{Rank Increasing Algorithm}
\label{rankinc}
In this section, we propose a rank-one update algorithm to generate good initial points and provide monotonic decrease for the objective functions for fixed-rank optimization in the  procedure of rank pursuit in Algorithm {\ref{rpcode}}.  This is achieved by exploiting the structure of the low-rank matrix varieties \cite{Schneider_2014convergencelinsearch,Vandereycken_NOLTA2014}.

\subsection{Low-Rank Matrix Varieties}
We present a systematic way to develop the rank increasing strategy in Algorithm {\ref{rpcode}} based on the  following low-rank matrix varieties
$\mathcal{M}_{\le r}=\{{\bf{X}}\in\mathbb{R}^{M\times M}: {\rm{rank}}({\bf{X}})\le
r\}$,
which is the closure of the set of fixed-rank metrics $\mathcal{M}_r$. Furthermore, we consider the linear-search method on $\mathcal{M}_{\le r+1}$ with
the iterates as follows,
\begin{eqnarray}
\label{lsear}
{\bf{X}}_{i+1}=P_{\le r+1}({\bf{X}}_{i}+\alpha_i {\bs{\Xi}}_{i}),
\end{eqnarray} 
where ${\bf{\Xi}}_i$ is a search direction in the \emph{tangent cone} $T_{{\bf{X}}_{i}}\mathcal{M}_{\le
r+1}$ at ${\bf{X}}_i$ \cite{Schneider_2014convergencelinsearch}, $\alpha_i$
is a step-size, and $P_{\le r+1}$ is a metric projection onto $\mathcal{M}_{\le
r+1}$ with a best rank-$(r+1)$ approximation in the Frobenius norm. 

\subsection{Riemannian Pursuit}
Assume that the iterate ${\bf{X}}^{[r]}$ has rank $r$ at the $r$-th iteration in
Algorithm {\ref{rpcode}}. In the next iteration, we will increase the rank
by $r+1$. 
To embed ${\bf{X}}^{[r]}$ into the search space $\mathcal{M}_{\le r+1}$,
suppose that we choose the  projection of the negative Euclidean gradient
on the tangent cone $T_{{\bf{X}}^{[r]}}\mathcal{M}_{\le r+1}$ as a search direction,
${\bf{\Xi}}_r=\mathop {\arg\min}_{{\bf{\Xi}}\in T_{{\bf{X}}^{[r]}}\mathcal{M}_{\le
r+1}}\|-\nabla_{{\bf{X}}^{[r]}} f-{\bs{\Xi}}\|_F={\bs{\Xi}}_r^{(r)}+{\bs{\Xi}}_r^{(1)}$,
where $\nabla_{{\bf{X}}^{[r]}} f=(\mathcal{P}_{\Omega}({\bf{X}}^{[r]})-{\bf{I}}_M)$ is the Euclidean gradient of the cost
function $f$ at point ${\bf{X}}^{[r]}$ and ${\bs{\Xi}}_r^{(r)}$ is the orthogonal projection on the tangent space
$T_{{\bf{X}}^{[r]}}\mathcal{M}_r$ given by the Riemannian gradient, i.e.,
${\bs{\Xi}}_r^{(r)}=-{\rm{grad}}_{{\bf{X}}^{[r]}}f$,
and ${\bs{\Xi}}_r^{(1)}$ is the best rank-one approximation of 
\begin{eqnarray}
\label{ru}
\!\!\!\!\!\!\!\!{\bs{\Sigma}}_r&=&-\nabla_{{\bf{X}}^{[r]}}
f-{\bs{\Xi}}_r^{(r)}-\nabla_{{\bf{X}}^{[r]}} f({\bf{X}}^{[r]})+{\rm{grad}}_{{\bf{X}}^{[r]}}
f\nonumber\\
&=&-\nabla_{{\bf{X}}^{[r]}} f({\bf{X}}^{[r]})+{\bs{\xi}}_{U}{\bs{\Sigma}}{\bf{V}}^T+{\bf{U}}{\bs{\xi}}_{{\Sigma}}{\bf{V}}^T+{\bf{U}}{\bf{\Sigma}}{\bs{\xi}}_{V}^T,
\end{eqnarray}
which is orthogonal to the tangent space $T_{{\bf{X}}^{[r]}}\mathcal{M}_{r}$
\cite{Absil2014low}.

Based on (\ref{lsear}) and (\ref{ru}), we shall adopt the following rank update strategy to
find a good initial point for the next iteration in Algorithm {\ref{rpcode}},
\begin{eqnarray}
\label{rpu}
{\bf{X}}_{0}^{[r+1]}=P_{\le r+1}\left({\bf{X}}^{[r]}+\alpha_r \left({\bs{\Xi}}_{r}^{(1)}-{\rm{grad}}_{{\bf{X}}^{[r]}}f\right)\right),
\end{eqnarray}
where $\alpha_r\ge0$ is a step size and satisfies the following condition \cite{Vandereycken_ICML2014RiemanMatrixRec},
\begin{eqnarray}
\label{ps}
f({\bf{X}}_{0}^{[r+1]})\le f({\bf{X}}^{[r]})-{{\alpha_{r}}\over{2}}\langle{\bf{\Theta}}_r,
{\bf{\Theta}}_r\rangle.
\end{eqnarray} 
Therefore, if ${\bf{\Xi}}_{r}$ is zero, then $\nabla_{{\bf{X}}^{[r]}}
f=0$ and we can terminate. 

\begin{remark}
Note that when the Riemannian gradient ${\rm{grad}}_{{\bf{X}}^{[r]}}f$ equals  zero,
the rank update strategy (\ref{rpu}) is equivalent to the following rank increasing strategy \cite{Mishra2013lowtracenorm} 
\begin{eqnarray}
\label{rankone}
{\bf{X}}_{0}^{[r+1]}={\bf{X}}^{[r]}-\sigma{\bf{u}}{\bf{v}}^{T},
\end{eqnarray} 
where $\sigma\ge0$ is the dominant singular value and $({\bf{u}}, {\bf{v}})$
is the pair of top left and right singular vectors with unit-norm of the
Euclidean gradient $\nabla_{{\bf{X}}^{[r]}} f$. Although the rank update
strategy
(\ref{rankone}) ensures that the cost function $f$ decreases monotonically
w.r.t. $r$, it ignores the intrinsic manifold structure of fixed-rank matrices
in Algorithm {\ref{manoptcode}}. Specifically, the Riemannian gradient ${\rm{grad}}_{{\bf{X}}^{[r]}}
f$
(\ref{gradrm}), which belongs to the tangent space $T_{{\bf{X}}^{[r]}}\mathcal{M}_{r}$,
is not necessarily equal to zero, as the corresponding fixed-rank optimization
problem may not be solved exactly in practice, e.g., Algorithm {\ref{manoptcode}}
may terminate when the maximum number of iterations is exceeded \cite{Vandereycken_ICML2014RiemanMatrixRec}.
\end{remark}

\subsection{Monotonic Decrease of the Objective Function}  We shall show
that
the Riemannian manifold rank update strategy (\ref{rpu}) ensures that the
objective function decreases monotonically with respect to $r$. Specifically,
as ${\rm{grad}}_{{\bf{X}}^{[r]}}f\in
T_{{\bf{X}}^{[r]}}\mathcal{M}_{r}$ and ${\bs{\Sigma}}_r$ (\ref{ru}) is orthogonal
to $T_{{\bf{X}}^{[r]}}\mathcal{M}_{r}$, we have the following fact that
\begin{eqnarray}
\label{ot}
\langle{\bs{\Sigma}}_r^{(1)}, {\rm{grad}}_{{\bf{X}}^{[r]}}
f\rangle=0.
\end{eqnarray}
Let ${\bf{X}}^{[1]},
{\bf{X}}^{[2]},\dots,$ be the sequence generated by Algorithm \ref{rpcode},
based on (\ref{ps}) and (\ref{ot}), we have 
\begin{eqnarray}
f({\bf{X}}^{[r+1]})&\le_{(1)}& f({\bf{X}}_{0}^{[r+1]})\le_{(2)} f({\bf{X}}^{[r]})-{{\alpha_{r}}\over{2}}\langle{\bf{\Theta}}_r,
{\bf{\Theta}}_r\rangle\nonumber\\
&\le_{(3)}& f({\bf{X}}^{[r]})-{{\tau_r}\over{2}}(\|{\bs{\Sigma}}_r^{(1)}\|_{F}^2+\!\|{\rm{grad}}_{{\bf{X}}^{[r]}}
f\|_F^2)\nonumber\\
&\le_{(4)}& f({\bf{X}}^{[r]}).
\end{eqnarray} 
Here, the first inequality is due to the fact that the iterates of the Riemannian
optimization algorithm try to minimize the cost function $f$, the second
and the third inequalities are based on the facts (\ref{ps}) and (\ref{ot}),
respectively. Therefore, the cost function $f({\bf{X}}^{[r]})$ decreases
monotonically with respect to $r$. 

\begin{remark}
Although only the rank-one update strategy is considered in Algorithm {\ref{rpcode}}, the proposed rank increasing algorithm in this section can be easily generalized to the general rank-$r$ with $r>1$ updates to improve the convergence rate \cite{Vandereycken_ICML2014RiemanMatrixRec,Vandereycken_NOLTA2014} for the RP algorithm. However, this may yield the  detected rank of matrix $\bf{X}$ overestimated.  
\end{remark}

\section{Simulation Results}
\label{simres}
In this section, we simulate the proposed Riemannian pursuit algorithms
for topological interference management problems in partially connected $K$-user interference
channels. The conjugate gradient Riemannian algorithm and the trust-region Riemannian pursuit algorithm, are  termed ``{{CGRP}}" and ``{{TRRP}}", respectively. The two algorithms are compared to  the following state-of-the-art algorithms:
\begin{itemize}
\item {{LRGeom with Riemannian Pursuit}}: In this algorithm \cite{Vandereycken_ICML2014RiemanMatrixRec, Yuanming_ISIT2015TIA}, termed ``LRGeom", the embedded manifold based fixed-rank optimization algorithm developed in \cite{Vandereycken2013low} with the Riemannian pursuit rank increasing strategy proposed in \cite{Yuanming_ISIT2015TIA,Vandereycken_ICML2014RiemanMatrixRec} is adopted to solve problem $\mathscr{P}$.
 \item {{LMaFit}}: In this algorithm, the alternating minimization scheme with rank adaptivity is adopted to solve problem $\mathscr{P}$ \cite{Wotao_2012solvingLR}.    
\end{itemize}

The Matlab implementation of all the Riemannian algorithms for the fixed-rank optimization problem $\mathscr{P}_r$ is based on the manifold optimization toolbox ManOpt \cite{manopt}. All the Riemannian optimization algorithms are initialized randomly as shown in \cite{Vandereycken2013low} and are terminated when either the norm of the Riemannian gradient is below $10^{-6}$, i.e., $\|{\rm{grad}}_{\bf{X}}f\|\le 10^{-6}$, or the number of iterations exceeds 500. The setting for {{LMaFit}} is the same as that in \cite{Wotao_2012solvingLR}. We adopt the following normalized residual \cite{Wotao_2012solvingLR} as the stopping criteria for Algorithm {\ref{rpcode}} to estimate the rank for matrix $\bf{X}$:
$\epsilon={{\|\mathcal{P}_{\Omega}({\bf{X}})-{\bf{I}}_M\|_F}/{\sqrt{M}}}$.
We set $\epsilon=10^{-6}$ for all the algorithms to estimate the minimum rank of matrix $\bf{X}$ such that it satisfies the affine constraint in problem $\mathscr{P}$. 
\subsection{Convergence Rate}
  \begin{figure}[t]
  \centering
  \includegraphics[width=0.95\columnwidth]{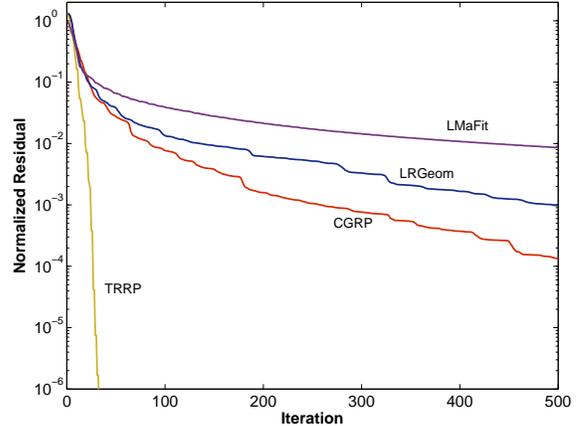}
 \caption{Convergence rate with the rank of matrix ${\bf{X}}$ as four.}
 \label{cong1}
 \end{figure}
  \begin{figure}[t]
  \centering
  \includegraphics[width=0.95\columnwidth]{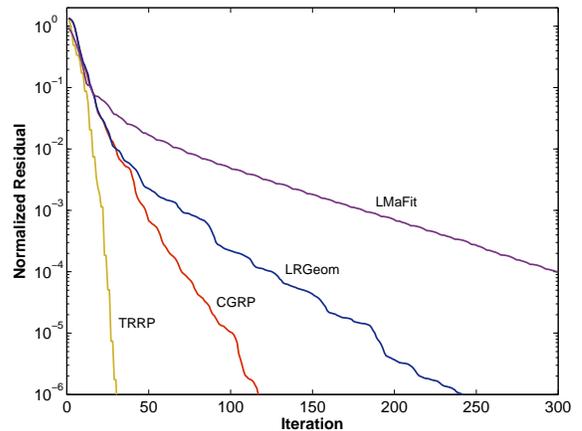}
 \caption{Convergence rate with the rank of matrix $\bf{X}$ as five.}
 \label{cong2}
 \end{figure}
 Consider a 100-user partially connected interference channel with 400 interference
channel links. The sets of the connected interference links  are generated uniformly at random. We turn off rank adaptivity for all the algorithms to solve the fixed-rank optimization problem $\mathscr{P}_r$. Fig. {\ref{cong1}} and Fig. {\ref{cong2}} show the convergence rates of different algorithms for the fixed-rank optimization problem $\mathscr{P}_r$ with $r=4$ and $r=5$, respectively. Both figures show that the trust-region based  Riemannian optimization algorithm TRRP has the fastest convergence rate and achieves higher precision solutions in a few iterations compared with the other three algorithms. Encoded with the second-order information in the Riemannian metric (\ref{rmetric}), the conjugate gradient based Riemannian algorithm CGRP achieves a faster convergence rate than {{LRGeom}} \cite{Vandereycken2013low}, while {{LMaFit}} \cite{Wotao_2012solvingLR} has the lowest convergence rate among all the algorithms.    

These two figures also indicate that, with the same stopping criteria $\epsilon=10^{-6}$ in Algorithm {\ref{rpcode}}, the detected rank of  matrix $\bf{X}$ by TRRP  is 4. Although the detected rank of matrix $\bf{X}$ by both CGRP  and LRGeom is 5,  the latter one has a slower convergence rate. Furthermore, the required rank of LMaFit should be larger than 5 to achieve the stopping criteria $\epsilon=10^{-6}$. This conclusion will be further confirmed in the following simulations on the empirical results for the achievable DoFs.

\subsection{Achievable Symmetric DoF and Optimal DoF Results}
  \begin{figure}[t]
  \centering
  \includegraphics[width=0.95\columnwidth]{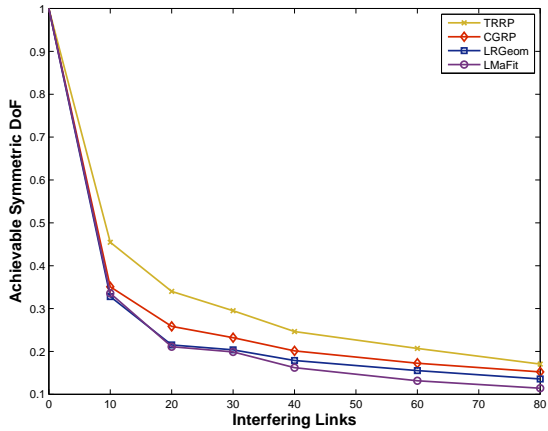}
 \caption{Achievable symmetric DoF versus different numbers of interference
links.}
 \label{sim1}
 \end{figure}
Consider a 20-user partially connected interference channel. The sets of the connected interference links  are generated
uniformly at random.
We simulate and average 100 network topology realizations. Fig. {\ref{sim1}} demonstrates the  achievable symmetric DoF with different algorithms assuming that the data streams $M_i=1, \forall i$. We can see that the second-order algorithm TRRP can achieve the highest symmetric DoF, but  it has the highest computational complexity due to the computation expensive calculation of the Hessian. For the first-order optimization algorithm, CGRP can achiever a higher symmetric DoF than  LRGeom \cite{Vandereycken_ICML2014RiemanMatrixRec,Yuanming_ISIT2015TIA} and LMaFit  \cite{Wotao_2012solvingLR}. In particular, we can see that, with few interference links, quite high DoFs can be achieved.  

{\rev{To further justify the effectiveness of the RP framework, we numerically
check that our RP algorithms can recover all the optimal DoF results for
the specific TIM problems in \cite{Jafar_TIT2013TIM}. The same conclusion
has also been presented
in \cite{Yuanming_ISIT2015TIA}. Note that our proposed automatic
rank detection capable RP algorithms do not need the optimal rank as a prior
information, while the alternating projection algorithm  \cite{Hassibi_TIA2014}
requires the optimal rank as a prior information to perform low-rank matrix
projection.
Moreover, it is interesting to theoretically identify the class of network
topologies
such that the proposed RP framework can provide optimal symmetric DoFs.}}

In summary, all the simulation results illustrate the effectiveness of the proposed Riemannian pursuit algorithms by exploiting the quotient manifold geometry of the fixed-rank matrices and encoding the second-order information in the Riemannian metric (\ref{rmetric}), as well as utilizing the second-order optimization scheme. {\rev{In particular, there is a tradeoff between the achievable symmetric
DoF and the computational complexity using the first-order algorithm CGRP
(which is applicable in large-sized networks) and the second-order algorithm
TRRP (which is applicable in small-sized and medium-sized networks).}}

\section{Conclusions and further works}
\label{confw}
In this paper, we presented a flexible low-rank matrix completion approach to maximize the achievable DoFs for the partially connected $K$-user interference channel with any network topology. A Riemannian pursuit algorithm was proposed to solve the resulting low-rank matrix completion optimization problem by exploiting the quotient manifold geometry of the search space   and the structure of low-rank matrix varieties for rank pursuit. In particular, we showed that, by encoding the second-order information, the quotient manifold based Riemannian optimization algorithms achieve a faster convergence rate and higher precious solutions than the existing algorithms. Simulation results showed that the proposed Riemannian pursuit algorithms achieve higher DoFs for general network topologies compared
with the state-of-the-art methods. 

Several future directions of interest are listed as follows:       
\begin{itemize}
\item From the algorithmic perspective, it is interesting to establish the optimality of the Riemannian pursuit algorithms for the low-rank matrix completion problem $\mathscr{P}$, thereby establishing the relationship between the achievable DoF and the network topology. 
\item From the information theoretic perspective, it is critical to translate the numerical insights {\rev{(e.g., optimal DoF achievability for the specific network topologies in \cite{Jafar_TIT2013TIM})}} provided by the LRMC approach into the optimal DoF for any network topology.
\item It is particularly interesting to extend the LRMC approach to more general scenarios, e.g., with finite SNR scenarios, MIMO interference channels, transmitter cooperations with data sharing, and  wired linear index coding problems in the finite field. {\rev{In particular, as  optimization on manifolds deeply relies on
smoothness,
the search space will become discrete in a finite field. Therefore, the presented
Riemannian pursuit algorithms cannot be extended to the finite field in principle.}} 
\item It is also interesting to apply the Riemannian optimization technique to other wireless communications and networking problems (e.g., the hybrid precoding in millimeter wave systems \cite{Letaief_JSTSP2016mmWave}). In particular, extending the corresponding algorithms to the complex field is critical, as most of the Riemannian algorithms are only developed in real field and complex field extension is not trivial. 
\end{itemize}

\appendices
\section{Proof of Proposition {\ref{proprm}}: Riemannian Metric}
\label{apprm}
To induce the metric based on the Hessian of the cost function $f$ in problem $\mathscr{P}_r$, we
consider a simplified cost function $\|{\bf{X}}-{\bf{I}}_M\|_F^2/2$, yielding the following optimization problem:
\begin{eqnarray}
\label{sfrank}
\mathop{\rm{minimize}}_{{\bf{X}}\in\mathcal{M}_r}~~{1\over{2}}{\rm{Tr}}({\bf{X}}^{T}{\bf{X}})-{\rm{Tr}}({\bf{X}}),
\end{eqnarray}
Based on the factorization ${\bf{X}}={\bf{U}}{\bf{\Sigma}}{\bf{V}}^T$, we have the matrix representation of Lagrangian for problem (\ref{sfrank}) as follows
$\mathcal{L}({\bf{X}})={1\over{2}}{\rm{Tr}}({\bf{V}}{\bf{\Sigma}}^T{\bf{U}}^T{\bf{U}}{\bf{\Sigma}}{\bf{V}}^T)-{\rm{Tr}}({\bf{U}}{\bf{\Sigma}}{\bf{V}}^T)$,
where ${\bf{X}}$ has the matrix representation $({\bf{U}}, {\bf{\Sigma}},
{\bf{V}})\in {\rm{St}}(r,n)\times
{\rm{GL}}(r)\times {\rm{St}}(r,n)$.
The second-order derivative of $\mathcal{L}(\bf{X})$ applied in the direction
${\bs{\xi}}_{\bf{X}}$ is given by
${D}^2\mathcal{L}({\bf{X}})[{\bs{\xi}}_{\bf{X}}]=({\bs{\xi}}_{\bf{U}}{\bf{\Sigma}}{\bf{\Sigma}}^T+2{\bf{U}}{\rm{Sym}}({\bf{\Sigma}}{\bs{\xi}}_{\bf{\Sigma}})-{\bf{V}}{\bs{\xi}}_{\bf{\Sigma}}-{\bs{\xi}}_{\bf{V}}{\bf{\Sigma}}^T,
-{\bs{\xi}}_{\bf{U}}{\bf{V}}^T+{\bs{\xi}}_{\bf{\Sigma}}+2{\bf{\Sigma}}{\rm{Sym}}({\bf{V}}^T{\bs{\xi}}_{\bf{V}})-{\bf{U}}^T{\bs{\xi}}_{\bf{V}},
{\bs{\xi}}_{\bf{V}}{\bf{\Sigma}}{\bf{\Sigma}}^T-{\bf{U}}{\bs{\xi}}_{\bf{\Sigma}}-{\bs{\xi}}_{\bf{U}}{\bf{\Sigma}}^T+2{\bf{V}}{\rm{Sym}}({\bf{\Sigma}}^T{\bs{\xi}}_{\bf{\Sigma}}))$,
where ${\bs{\xi}}_{\bf{X}}$ has the matrix
representation $({\bs{\xi}}_{\bf{U}}, {\bs{\xi}}_{\bf{\Sigma}}, {\bs{\xi}}_{\bf{V}})\in\mathbb{R}^{n\times
r}\times \mathbb{R}^{r\times r}\times \mathbb{R}^{n\times r}$.

As the cost function in (\ref{sfrank}) is convex and quadratic in $\bf{X}$,
it is also convex and quadratic in the arguments $({\bf{U}}, {\bf{\Sigma}},
{\bf{V}})$ individually. Therefore, the block diagonal elements of the second-order
derivative $\mathcal{L}_{\bf{XX}}({\bf{X}})$ of the Lagrangian are strictly
positive definite. The following Riemannian metric can be induced from the
block diagonal approximation of $\mathcal{L}_{\bf{XX}}(\bf{X})$,
\begin{eqnarray}
\label{metricrm}
g_{\bf{X}}({\bs{\xi}}_{\bf{X}}, {\bs{\zeta}}_{\bf{X}})&=&\langle {\bs{\xi}}_{\bf{X}},
{D}^2\mathcal{L}({\bf{X}})[{\bs{\zeta}}_{\bf{X}}]\rangle\nonumber\\
&\approx&\langle {\bs{\xi}}_{\bf{U}}, {\bs{\zeta}}_{\bf{U}}{\bf{\Sigma}}{\bf{\Sigma}}^T\rangle+\langle
{\bs{\xi}}_{\bf{\Sigma}}, {\bs{\zeta}}_{\bf{\Sigma}}\rangle+\nonumber\\
&&\langle {\bs{\xi}}_{\bf{V}},{\bs{\zeta}}_{\bf{V}}{\bf{\Sigma}}^T{\bf{\Sigma}}\rangle,
\end{eqnarray} 
where ${\bs{\xi}}_{\bf{X}}=({\bs{\xi}}_{\bf{U}}, {\bs{\xi}}_{\bf{\Sigma}},
{\bs{\xi}}_{\bf{V}}), {\bs{\zeta}}_{\bf{X}}=({\bs{\zeta}}_{\bf{U}}, {\bs{\zeta}}_{\bf{\Sigma}},
{\bs{\zeta}}_{\bf{V}})\in T_{\bf{X}}\mathcal{M}_r$ and ${\bf{X}}\in({\bf{U}},
{\bf{\Sigma}}, {\bf{V}})$. 

To verify that the metric is invariant along the equivalent class $[\bf{X}]$ (\ref{ecla}), based
on \cite[Proposition 3.6.1]{Absil_2009optimizationonManifolds}, it is equivalent
to show that the metric for tangent vectors ${\bs{\xi}}_{\bf{X}}, {\bs{\zeta}}_{\bf{X}}\in
T_{\bf{X}}\mathcal{M}_r$ does not change under the transformations $({\bf{U}},
{\bf{\Sigma}}, {\bf{V}})\mapsto ({\bf{U}}{\bf{Q}}_U, {\bf{Q}}_U^T{\bf{\Sigma}}{\bf{Q}}_V,
{\bf{Q}}_V{\bf{V}})$, $({\bs{\xi}}_U, {\bs{\xi}}_{{\Sigma}}, {\bs{\xi}}_{{V}})\mapsto
({\bs{\xi}}_U{\bf{Q}}_U,
{\bf{Q}}_U^T{\bs{\xi}}_{{\Sigma}}{\bf{Q}}_V, {\bs{\xi}}_{{V}}{\bf{V}})$,
 $({\bs{\zeta}}_U,
{\bs{\zeta}}_{{\Sigma}}, {\bs{\zeta}}_{{V}})\mapsto ({\bs{\zeta}}_U{\bf{Q}}_U,
{\bf{Q}}_U^T{\bs{\zeta}}_{{\Sigma}}{\bf{Q}}_V, {\bs{\zeta}}_{{V}}{\bf{V}})$. After simple computation, we can verify that (\ref{metricrm}) is a valid Riemannian metric and does not depend on the specific matrix representations along the equivalence class $[\bf{X}]$ (\ref{ecla}).

\section {Proof of Proposition \ref{prohs}: Horizontal Space}
\label{apphs}

The {vertical space} $\mathcal{V}_{{\bf{X}}}{\mathcal{M}}_r$ is the linearization of  the equivalence classes $[{\bf{X}}]$ (\ref{ecla}) and formed by the set of directions that contains tangent vectors to the equivalence classes. Based on the matrix representation of the tangent space for the orthogonal matrices \cite[Example 3.5.3]{Absil_2009optimizationonManifolds}, we have the matrix representation for the vertical space as 
\begin{eqnarray}
\mathcal{V}_{{\bf{X}}}{\mathcal{M}}_r=({\bf{U}}{\bf{\Theta}}_1, {\bs{\Sigma}}{\bf{\Theta}}_2-{\bf{\Theta}}_1{\bf{\Sigma}},
{\bf{V}}{\bf{\Theta}}_2),
\end{eqnarray} 
where ${\bf{\Theta}}_1$ and ${\bf{\Theta}}_2$ are any skew-symmetric matrices of size
$r\times r$, i.e., ${\bf{\Theta}}_i^T=-{\bf{\Theta}}_i, i=1,2$.

The {horizontal space} $\mathcal{H}_{\bf{X}}\mathcal{M}_r$,
which is any complementary subspace to $\mathcal{V}_{\bf{X}}\mathcal{M}_r$ in $T_{\bf{X}}\mathcal{M}_r$
with respect to the {Riemannian metric} $g_{\bf{X}}$ (\ref{rmetric}), provides a valid matrix
representation of the abstract tangent space $T_{[\bf{X}]}(\mathcal{M}_r/\sim)$
\cite[Section 3.5.8]{Absil_2009optimizationonManifolds} based on the Riemannian submersion principle. Specifically, let ${\bs{\eta}}_{\bf{X}}=({\bs{\eta}}_{\bf{U}}, {\bs{\eta}}_{\bf{\Sigma}},
{\bs{\eta}}_{\bf{V}})\in {\mathcal{H}}_{\bf{X}}\mathcal{M}_r$ and ${\bs{\zeta}}_{\bf{X}}=({\bs{\zeta}}_{\bf{U}}, {\bs{\zeta}}_{\bf{\Sigma}}, {\bs{\zeta}}_{\bf{V}})\in\mathcal{V}_{\bf{X}}\mathcal{M}_r$. By definition,
${\bs{\eta}}_{\bf{X}}$ should be orthogonal to ${\bs{\zeta}}_{\bf{X}}$ with respect to the Riemannian metric $g_{\bf{X}}$,
i.e.,
\begin{eqnarray}
g_{\bf{X}}({\bs{\eta}}_{\bf{X}}, {\bs{\zeta}}_{\bf{X}})&=&{\rm{Tr}}(({\bf{\Sigma}}{\bf{\Sigma}}^{T}){\bs{\eta}}_{\bf{U}}^{T}{\bf{U}}{\bf{\Theta}}_1)+\nonumber\\
&&{\rm{Tr}}({\bs{\eta}}_{\bf{\Sigma}}^{T}{\bs{\Sigma}}{\bf{\Theta}}_2-{\bs{\eta}}_{\bf{\Sigma}}^{T}{\bf{\Theta}}_1{\bf{\Sigma}})+\nonumber\\
&&{\rm{Tr}}(({\bf{\Sigma}}^T{\bf{\Sigma}}){\bs{\eta}}_{\bf{V}}^{T}{\bf{V}}{\bf{\Theta}}_2)\nonumber\\
&=& {\rm{Tr}}({\bf{S}}_1{\bf{\Theta}}_1)+{\rm{Tr}}({\bf{S}}_2{\bf{\Theta}}_2)=0,
\end{eqnarray} 
where ${\bf{S}}_1={\bf{\Sigma}}{\bf{\Sigma}}^{T}{\bs{\eta}}_{\bf{U}}^{T}{\bf{U}}-{\bf{\Sigma}}{\bs{\eta}}_{\bf{\Sigma}}^T$
and ${\bf{S}}_2={\bf{\Sigma}}^{T}{\bf{\Sigma}}{\bs{\eta}}_{\bf{V}}^T{\bf{V}}+{\bs{\eta}}_{\bf{\Sigma}}^T{\bf{\Sigma}}$.
Based on the fact that ${\rm{Tr}}({\bf{G}}^T{\bf{\Theta}})={\bf{0}}$,
if and only if ${\bf{G}}$ is symmetric, the characterization of the horizontal space is given by
\begin{eqnarray}
\label{hspace}
\!\!\!\!\!\!\mathcal{H}_{\bf{X}}\mathcal{M}_r=\{{\bs{\eta}}_{\bf{X}}\in
T_{\bf{X}}\mathcal{M}_r: {\bf{S}}_1~{\rm{and}}~{\bf{S}}_2~{\rm{are
~symmetric}}\}.
\end{eqnarray}

\section{Proof of Proposition {\ref{protp}}: Tangent Space Projection}
\label{apptp}
Given a matrix in the ambient space $\mathbb{R}^{M\times r}\times\mathbb{R}^{r\times r}\times \mathbb{R}^{M\times r}$, its projection onto the tangent space $T_{\bf{X}}\mathcal{M}_r$ is obtained by extracting the component normal space $N_{\bf{X}}\mathcal{M}_r$ to the tangent space in the Riemannian metric sense. 

We first derive the matrix characterization of the normal space. Specifically, let ${\bs{\eta}}_{\bf{X}}=({\bs{\eta}}_{\bf{U}},
{\bs{\eta}}_{\bf{\Sigma}},
{\bs{\eta}}_{\bf{V}})\in T_{\bf{X}}\mathcal{M}_r$ and ${\bs{\zeta}}_{\bf{X}}=({\bs{\zeta}}_{\bf{U}},
{\bs{\zeta}}_{\bf{\Sigma}}, {\bs{\zeta}}_{\bf{V}})\in N_{\bf{X}}\mathcal{M}_r$.
By definition,
${\bs{\eta}}_{\bf{X}}$ should be orthogonal to ${\bs{\zeta}}_{\bf{X}}$ with
respect to the Riemannian metric $g_{\bf{X}}$,
i.e., $g({\bs{\eta}}_{\bf{X}}, {\bs{\zeta}}_{\bf{X}})=0$. That is, the  following conditions
\begin{eqnarray}
\label{normc1}
\!\!\!\!\langle {\bs{\xi}}_{\bf{U}}, {\bs{\zeta}}_{\bf{U}}{\bf{\Sigma}}{\bf{\Sigma}}^T\rangle=0,\langle {\bs{\xi}}_{\bf{V}},
{\bs{\zeta}}_{\bf{V}}{\bf{\Sigma}}^T{\bf{\Sigma}}\rangle=0, \langle
{\bs{\xi}}_{\bf{\Sigma}}, {\bs{\zeta}}_{\bf{\Sigma}}\rangle=0,
\end{eqnarray} 
should hold for any ${\bs{\eta}}_{\bf{X}}\in T_{\bf{X}}\mathcal{M}_r$. It is obvious that ${\bs{\zeta}}_{\bf{\Sigma}}={\bf{0}}$. Furthermore, based on \cite[Example 3.5.2]{Absil_2009optimizationonManifolds}, we have the matrix characterization of ${\bs{\eta}}_{\bf{U}}$ as
\begin{eqnarray}
\label{normc2}
{\bs{\eta}}_{\bf{U}}={\bf{U}}{\bf{\Omega}}+{\bf{U}}_{\bot}{\bf{K}},
\end{eqnarray}
where $\bf{\Omega}$ is a skew-symmetric matrix of size $r\times r$, ${\bf{K}}\in \mathbb{R}^{(M-r)\times r}$ can be any matrix, and ${\bf{U}}_{\bot}$ is any $M\times (M-r)$ matrix such that ${\rm{span}}({\bf{X}}_{\bot})$ is the orthogonal complement of ${\rm{span}}(\bf{X})$. Similarly, we can obtain the characterization of ${\bs{\eta}}_{\bf{V}}$. We rewrite ${\bs{\zeta}}_{\bf{U}}$ as $\bar{\bs{\zeta}}_{\bf{U}}={\bs{\zeta}}_{\bf{U}}{\bf{\Sigma}}{\bf{\Sigma}}^T$ with,
\begin{eqnarray}
\bar{\bs{\zeta}}_{\bf{U}}={\bf{U}}{\bf{B}}_U+{\bf{U}}_{\bot}{\bf{A}}_U,
\end{eqnarray}
where ${\bf{A}}_U\in\mathbb{R}^{r\times r}$ and ${\bf{B}}_U\in\mathbb{R}^{(M-r)\times r}$ can be deduced from conditions (\ref{normc1}) and (\ref{normc2}). Based on the fact that ${\rm{Tr}}({\bf{G}}^T{\bf{\Theta}})={\bf{0}}$,
if and only if ${\bf{G}}$ is symmetric, we can conclude that ${\bf{B}}_U$ is symmetric and ${\bf{A}}_U=\bf{0}$. Therefore, we have
\begin{eqnarray}
{\bs{\zeta}}_{\bf{U}}{\bf{\Sigma}}{\bf{\Sigma}}^T={\bf{U}}{\bf{B}}_U,
\end{eqnarray}
where ${\bf{B}}_U={\bf{B}}_U^T$. Similarly, we can obtain the matrix characterization of ${\bs{\zeta}}_{\bf{V}}$. Therefore, we arrive at the matrix representation of the norm space,
\begin{eqnarray}
N_{\bf{X}}\mathcal{M}_r=\{({\bf{U}}{\bf{B}}_U({\bf{\Sigma}}{\bf{\Sigma}}^T)^{-1}, {\bf{0}}, {\bf{V}}{\bf{B}}_V({\bf{\Sigma}}^T{\bf{\Sigma}})^{-1})\},
\end{eqnarray}          
where ${\bf{B}}_U$ and ${\bf{B}}_V$ are symmetric metrics of size $r\times r$.

As the tangent space projector $P_{T_{\bf{X}}}\mathcal{M}_r$ is obtained by extracting the component normal to the tangent space $T_{\bf{X}}\mathcal{M}_r$ in the ambient space $\mathbb{R}^{M\times r}\times \mathbb{R}^{r\times r}\times \mathbb{R}^{M\times r}$, we have the expression for the operator $P_{T_{\bf{X}}}\mathcal{M}_r$ as
\begin{eqnarray}
\label{tproj}
\!\!\!\!\!\!\!\!P_{T_{\bf{X}}}\mathcal{M}_r({\bf{A}}_U, {\bf{A}}_{\Sigma}, {\bf{A}}_{V})&=&({\bf{A}}_{U}-{\bf{U}}{\bf{B}}_U({\bf{\Sigma}}{\bf{\Sigma}}^T)^{-1},\nonumber\\ &&{\bf{A}}_{\Sigma}, {\bf{A}}_V-{\bf{V}}{\bf{B}}_V({\bf{\Sigma}}^T{\bf{\Sigma}})^{-1})),
\end{eqnarray}
which belongs to the tangent space. The tangent space $T_{\bf{X}}\mathcal{M}_r$ in the computation space $\mathcal{M}_r$
at the point ${\bf{X}}=({\bf{U}}, {\bf{\Sigma}}, {\bf{V}})$ is the product
 of the tangent spaces of the individual manifolds, which has the following
matrix representation \cite[Example 3.5.2]{Absil_2009optimizationonManifolds},
\begin{eqnarray}
\label{matang}
T_{\bf{X}}\mathcal{M}_r&=&\{({\bs{\xi}}_{{U}}, {\bs{\xi}}_{{{\Sigma}}},
{\bs{\xi}}_{{V}})\in\mathbb{R}^{M\times r}\times\mathbb{R}^{r\times r}\times
\mathbb{R}^{M\times r}:\nonumber\\
&&{\bf{U}}^{T}{\bs{\xi}}_{{U}}+{\bs{\xi}}_{{U}}^T{\bf{U}}={\bf{0}}, {\bf{V}}^{T}{\bs{\xi}}_{{V}}+{\bs{\xi}}_{{V}}^T{\bf{V}}={\bf{0}}\}.
\end{eqnarray} 
Based on (\ref{tproj}) and (\ref{matang}), we know that ${\bf{U}}$ should satisfy the  condition:
\begin{eqnarray}
{\bf{U}}^{T}{\bs{\xi}}_{{U}}+{\bs{\xi}}_{{U}}^T{\bf{U}}&=&{\bf{U}}^T\left[{\bf{A}}_{U}-{\bf{U}}{\bf{B}}_U({\bf{\Sigma}}{\bf{\Sigma}}^T)^{-1}\right]+\nonumber\\
&&\left[{\bf{A}}_{U}-{\bf{U}}{\bf{B}}_U({\bf{\Sigma}}{\bf{\Sigma}}^T)^{-1}\right]^T{\bf{U}}={\bf{0}},
\end{eqnarray}
which is equivalent to
the  Lyapunov equation for the symmetric matrix ${\bf{B}}_U$,
\begin{eqnarray}
{\bf{\Sigma}}{\bf{\Sigma}}^{T}{\bf{B}}_{{U}}+{\bf{B}}_{{U}}{\bf{\Sigma}}{\bf{\Sigma}}^{T}={\bf{\Sigma}}{\bf{\Sigma}}^{T}({\bf{U}}^T{\bf{A}}_{{U}}+{\bf{A}}_{{U}}^{T}{\bf{U}}){\bf{\Sigma}}{\bf{\Sigma}}^{T}.
\end{eqnarray}
Similarly, we can obtain the Lyapunov equation for the symmetric matrix ${\bf{B}}_V$ as in (\ref{lyapeq1}).

\section{Proof of Proposition {\ref{prohp}}: Horizontal Space Projection}
\label{apphp}
The horizontal space projector $\Pi_{\mathcal{H}_{\bf{X}}\mathcal{M}_r}$ can be obtained by extracting the horizontal component of the tangent vector. Specifically, let ${\bs{\xi}}_{\bf{X}}=({\bs{\xi}}_{{U}},
{\bs{\xi}}_{{\Sigma}},
{\bs{\xi}}_{{V}})\in T_{\bf{X}}\mathcal{M}_r$ and ${\bs{\zeta}}_{\bf{X}}=({\bs{\zeta}}_{{U}},
{\bs{\zeta}}_{{\Sigma}}, {\bs{\zeta}}_{{V}})\in\mathcal{H}_{\bf{X}}\mathcal{M}_r$. We have the expression for the operator $\Pi_{\mathcal{H}_{\bf{X}}\mathcal{M}_r}$ as
\begin{eqnarray}
\Pi_{\mathcal{H}_{\bf{X}}\mathcal{M}_r}({\bs{\xi}}_{\bf{X}})&=&({\bs{\xi}}_{{U}}-{\bf{U}}{\bf{\Theta}}_1, {\bs{\xi}}_{{\Sigma}}+{\bf{\Theta}}_1{\bf{\Sigma}}-{\bf{\Sigma}}{\bf{\Theta}}_2,\nonumber\\
&&{\bs{\xi}}_{{V}}-{\bf{V}}{\bf{\Theta}}_2)\nonumber\\
&=&({\bs{\zeta}}_{{U}},
{\bs{\zeta}}_{{\Sigma}}, {\bs{\zeta}}_{{V}}),
\end{eqnarray}
which belongs to the horizontal space $\mathcal{H}_{\bf{X}}\mathcal{M}_r$. Based on (\ref{hspace}), we have
\begin{eqnarray}
{\bf{\Sigma}}{\bf{\Sigma}}^{T}{\bs{\zeta}}_{{U}}^{T}{\bf{U}}-{\bf{\Sigma}}{\bs{\zeta}}_{{\Sigma}}^T&=&{\bf{\Sigma}}{\bf{\Sigma}}^{T}{({\bs{\xi}}_{{U}}-{\bf{U}}{\bf{\Theta}}_1)}^{T}{\bf{U}}-\nonumber\\
&&{\bf{\Sigma}}{({\bs{\xi}}_{{\Sigma}}+{\bf{\Theta}}_1{\bf{\Sigma}}-{\bf{\Sigma}}{\bf{\Theta}}_2)}^T\nonumber\\
&=& ({\bf{\Sigma}}{\bf{\Sigma}}^{T}{\bs{\xi}}_U^T{\bf{U}}-{\bf{\Sigma}}{\bs{\xi}}_{\Sigma}^T)+({\bf{\Sigma}}{\bf{\Sigma}}^{T}{\bf{\Theta}}_1+\nonumber\\
&&{\bf{\Sigma}}{\bf{\Sigma}}^{T}{\bf{\Theta}}_1-{\bf{\Sigma}}{\bf{\Theta}}_2{\bf{\Sigma}}^T),
\end{eqnarray}
which is symmetric.  As  ${\bf{\Sigma}}{\bf{\Sigma}}^{T}{\bs{\zeta}}_{{U}}^{T}{\bf{U}}-{\bf{\Sigma}}{\bs{\zeta}}_{{\Sigma}}^T=({\bf{\Sigma}}{\bf{\Sigma}}^{T}{\bs{\zeta}}_{{U}}^{T}{\bf{U}}-{\bf{\Sigma}}{\bs{\zeta}}_{{\Sigma}}^T)^T$, we can obtain the equation in (\ref{coupledly1}). Similarly, we can obtain the equation in (\ref{coupledly2}) by checking the condition that ${\bs{\zeta}}_{V}$ is symmetric. 
  
 \section{Compute the Riemannian Gradient (\ref{gradrm})}
 \label{gradrmapp}
 Let ${\bf{X}}=({\bf{U}}, {\bf{\Sigma}}, {\bf{V}})$ and ${\bf{A}}=\nabla f({\bf{X}})=\mathcal{P}_{\Omega}({\bf{X}})-{\bf{I}}$ denote the Euclidean gradient of $f$ at point $\bf{X}$. The partial derivatives of $f({\bf{X}})$ with respective to ${\bf{U}}, {\bf{\Sigma}}$ and ${\bf{V}}$ are given by
\begin{eqnarray}
\!\!\!\!\!\!\!\!{{\partial f({\bf{X}})}\over{\partial{\bf{U}}}}={\bf{A}}{\bf{V}}{\bf{\Sigma}}^T, {{\partial f({\bf{X}})}\over{\partial{\bf{\Sigma}}}}={\bf{U}}^T{\bf{A}}{\bf{V}}, {{\partial f({\bf{X}})}\over{\partial{\bf{V}}}}={\bf{A}}^T{\bf{U}}{\bf{\Sigma}}.
\end{eqnarray}
With metric (\ref{rmetric}), the scaled Euclidean gradient  is given by
\begin{eqnarray}
\bar{\bf{A}}=({\bf{A}}{\bf{V}}{\bf{\Sigma}}^T({\bf{\Sigma}}{\bf{\Sigma}}^T)^{-1}, {\bf{U}}^T{\bf{A}}{\bf{V}}, {\bf{A}}^T{\bf{U}}{\bf{\Sigma}}({\bf{\Sigma}}^T{\bf{\Sigma}})^{-1}).
\end{eqnarray}
By further projecting $\bar{\bf{A}}$ onto the tangent space based on (\ref{tanpro}), we have the matrix representation (horizontal lift) ${\rm{grad}}_{\bf{X}}f$ of ${\rm{grad}}_{[\bf{X}]}f$ as
\begin{eqnarray}
{\rm{grad}}_{\bf{X}}f=P_{T_{\bf{X}}\mathcal{M}_r}(\bar{\bf{A}}),
\end{eqnarray}
which yields the equations in (\ref{gradrm}). Note that, based
on the Riemannian submersion principle \cite[Section 3.6]{Absil_2009optimizationonManifolds},
$P_{T_{\bf{X}}\mathcal{M}_r}(\bar{\bf{A}})$ is already the horizontal
lift, which can be verified that the horizontal space projection $\Pi_{\mathcal{H}_{{\bf{X}}}{\mathcal{M}}_r}$
will not change $P_{T_{\bf{X}}\mathcal{M}_r}(\bar{\bf{A}})$.

\section{Riemannian quotient manifolds}
We now consider the case of a quotient manifold ${\mathcal{M}}/\sim$, where the structure space $\mathcal{M}$ is endowed with a Riemannian metric $g$. The horizontal space $\mathcal{H}_{\bf{X}}$ and ${\bf{X}}\in\mathcal{M}$ is canonically chosen as the orthogonal complement in $T_{\bf{X}}\mathcal{M}$ of the vertical space $\mathcal{V}_{\bf{X}}=T_{\bf{X}}\pi^{-1}({\bf{X}})$, namely, 
\begin{eqnarray}
\mathcal{H}_{\bf{X}}&:=&(T_{\bf{X}}\mathcal{V}_{\bf{X}})^\perp\nonumber\\
&=&\{{\bs{\eta}}_{\bf{X}}\in T_{\bf{X}}\mathcal{M}:g({\bs{\chi}}_{\bf{X}}, {\bs{\eta}}_{\bf{X}})=0, \forall {\bs{\chi}}_{\bf{X}}\in\mathcal{V}_{\bf{X}}\}.
\end{eqnarray} 

Recall that the horizontal lift at ${\bf{X}}\in\pi^{-1}({[\bf{X}]})$ of a tangent vector ${\bs{\xi}}_{[\bf{X}]}\in T_{[\bf{X}]}(\mathcal{M}/\sim)$ is the unique tangent vector ${\bs{\xi}}_{\bf{X}}\in\mathcal{H}_{\bf{X}}$ that satisfies $D\pi({\bf{X}})[{\bs{\xi}}_{\bf{X}}]$. If, for every $[{\bf{X}}]\in\mathcal{M}/\sim$ and every ${\bs{\xi}}_{[\bf{X}]}, {\bs{\zeta}}_{[\bf{X}]}\in T_{[\bf{X}]}(\mathcal{M}/\sim)$, the expression $g_{\bf{X}}({\bs{\xi}}_{\bf{X}}, {\bs{\zeta}}_{\bf{X}})$ does not depend on ${\bf{X}}\in\pi^{-1}([\bf{X}])$, then
\begin{eqnarray}
g_{[\bf{X}]}({\bs{\xi}}_{[\bf{X}]}, {[\bs{\zeta}]}_{\bf{X}}):=g_{\bf{X}}({\bs{\xi}}_{\bf{X}}, {\bs{\zeta}}_{\bf{X}})
\end{eqnarray}
defines a Riemannian metric on $\mathcal{M}/\sim$. Endowed with this Riemannian metric, $\mathcal{M}/\sim$ is called a \emph{Riemannian quotient manifold} of $\mathcal{M}$, and the natural projection $\pi: \mathcal{M}\rightarrow \mathcal{M}/\sim$ is a \emph{Riemannian submersion}.  (In other words, a Riemannian submersion is a submersion of Riemannian manifolds such that $D\pi$ preserves inner products of vectors normal to fibers.)

\section*{Acknowledgment}
The authors would like to thank Dr. Bamdev Mishra, Dr. Nicolas Boumal and Prof. Bart Vandereycken for insightful discussions about Riemannian optimization for low-rank matrix completion.

\end{document}